\documentstyle[preprint,aps,epsf,tighten]{revtex}  

\newcommand{\slashed}[1]{\rlap{$#1$}/}
\newcommand{\slashp}{\mbox{$\not \hspace*{-1.10mm} p$}}

\newcommand{\slashq}{\mbox{$\not \hspace*{-1.10mm} q$}}

\newcommand{\GeV}{\mbox{\rm GeV}}

\begin{document}

\draft

\preprint{Zagreb University preprint ZTF-98/03}

\title{
Application of Jain and Munczek's bound--state \\ approach 
to $\gamma\gamma$--processes of $\pi^0, \eta_c$ and $\eta_b$
}

\author{Dalibor Kekez}
\address{\footnotesize Rudjer Bo\v{s}kovi\'{c} Institute,
         P.O.B. 1016, 10001 Zagreb, Croatia}
\author{Bojan Bistrovi\' c and Dubravko Klabu\v{c}ar}
\address{\footnotesize Department of Physics, Faculty of Science, \\
        Zagreb University, P.O.B. 162, 10001 Zagreb, Croatia}

\maketitle

\begin{abstract}
We point out the problems affecting most quark--antiquark bound state
approaches when they are faced with the electromagnetic processes 
dominated by Abelian axial anomaly. However, these problems are resolved 
in the consistently coupled Schwinger-Dyson and Bethe-Salpeter approach. 
Using one of the most successful variants of this approach, we find the 
dynamically dressed propagators of the light $u$ and $d$ quarks, as well 
as the heavy $c$ and $b$ quarks, and find the Bethe-Salpeter amplitudes 
for their bound states $\pi^0, \eta_c$ and $\eta_b$.
Thanks to incorporating the dynamical chiral symmetry breaking,
the pion simultaneously appears as the (pseudo)Goldstone boson. 
We give the theoretical predictions for the $\gamma\gamma$ decay 
widths of $\pi^0, \eta_c$ and $\eta_b$, and for
the $\pi^0\gamma^\star\to\gamma$ transition form factor,
and compare them with experiment.
In the chiral limit, the axial-anomaly result for $\pi^0\to\gamma\gamma$ 
is reproduced analytically in the consistently coupled Schwinger-Dyson and 
Bethe--Salpeter approach, provided that the quark-photon vertex is dressed
consistently with the quark propagator, so that the vector Ward-Takahashi 
identity of QED is obeyed.  On the other hand, the present approach is also 
capable of quantitatively describing systems of heavy quarks, concretely 
$\eta_c$ and possibly $\eta_b$, and their $\gamma\gamma$-decays. 
We discuss the reasons for the broad phenomenological success of the 
bound--state approach of Jain and Munczek. 

\vspace{1mm}

\indent \indent Electromagnetic processes; Constituent quarks; Current quarks
\end{abstract}
\renewcommand{\thefootnote}{\alph{footnote}}

\section{Introduction 
}
\label{INTRO-MOT}

\noindent

In Ref. \cite{KeKl1}, we calculated the two--photon decays
$\pi^0, \eta_c, \eta_b\rightarrow \gamma\gamma$, and 
-- in the case of the neutral pion -- also the off-shell
extension thereof, namely the transition form factor
for the process $\gamma^\star\pi^0 \rightarrow \gamma$.
We used a kind of quark--antiquark ($q\bar q$) bound-state 
approach to mesons known as the coupled Schwinger-Dyson and 
Bethe-Salpeter (SD-BS) approach.
More specifically, we used one of the phenomenologically most 
successful variant of this approach so far, namely the 
one of Jain and Munczek \cite{jain91,munczek92,jain93b}.

One of the motivations of this paper is to give the details
of these calculations which could not be presented in the 
letter \cite{KeKl1} for reasons of space, such as 
at least some of the solutions obtained with 
the gluon propagator {\it Ansatz} used by us,
and the comparison of some other gluon propagator 
{\it Ans\" atze} with this gluon propagator used by us. 

Another motivation is to present an updated comparison with 
the $\eta_c$ data, in order to show that the conclusions 
of \cite{KeKl1} are still valid in spite of some of the 
recent developments regarding the measurements of
$\eta_c\to\gamma\gamma$, which -- superficially -- may 
seem to have worked against the conclusions of \cite{KeKl1} 
regarding $\eta_c$. 

The third motivation lies in presenting the results for the
$\gamma\gamma$-decay of the massive $\pi^0$, since in Ref. 
\cite{KeKl1} the pion was treated in the chiral limit.
Although the difference between the ``massive" and the 
earlier, chiral-limit results for $\pi^0\to\gamma\gamma$
are (predictably) small, we discuss them because of their
importance as a test of validity of some recent attempts 
to take the SD-BS approach beyond ladder approximation.
These results are therefore also tied to the last, fourth
motivation.

Namely, the fourth motivation -- which is the most important one --
is to present more detailed explanations and discussions of 
why the coupled SD-BS approach can be so successful over 
such a surprisingly broad range of masses, and why this 
type of description of electromagnetic processes of mesons
in the coupled SD-BS approach, represents a significant advance
in understanding of mesons as $q\bar q$ bound states.
In brief, this is because this approach incorporates 
the dynamical chiral symmetry breaking (D$\chi$SB) into the 
bound states consistently, so that the pion,
although constructed as a $q\bar q$ bound state, appears as a
Goldstone boson in the chiral limit. Precisely because of this
feature, the coupled SD-BS approach correctly incorporates the 
Abelian axial (or Adler-Bell-Jackiw \cite{Adler69,BellJackiw69}, 
ABJ) anomaly of QED, which other bound-state approaches -- the ones
without D$\chi$SB -- simply cannot. These latter approaches
consequently fail to provide adequate descriptions of the
processes where the axial anomaly is important,
such as the two-photon processes of light pseudoscalars, of
which the $\pi^0\to\gamma\gamma$ decay is the cleanest
example of an anomalous electromagnetic decay. It is
important to elucidate this issue in detail,
because it is very often simply swept under the carpet
in various model approaches to such processes -- especially 
in the bound-state calculations. For example, model results 
for $\eta\to\gamma\gamma$ and $\eta^\prime\to\gamma\gamma$
which may seem reasonable by themselves, cannot be really 
physically significant if the analogous calculation of the 
closely related decay $\pi^0\to\gamma\gamma$ gives the result 
wrong by orders of magnitude. Similarly, if some approach yields  
good results for the {\it normalized} transition form factors 
for $\gamma^\star\pi^0 (\eta,\eta^\prime) \rightarrow \gamma$,
they are only seemingly good if the results for the related 
on-shell processes $\pi^0 (\eta,\eta^\prime) \to\gamma\gamma$ 
are wrong -- since this of course
implies equally wrong results for the absolute, un-normalized rates
for $\gamma^\star\pi^0 (\eta,\eta^\prime) \rightarrow \gamma$ at 
{\it any} transferred momentum. (See Eq. (\ref{transFF}) below.)

Therefore, in the next section we present a survey of
the situation in bound-state approaches concerning the
axial anomaly. In the third section we introduce the
coupled SD-BS approach, and in the fourth, we sketch how
the $q\bar q$ bound-state solutions are obtained. In the
fifth section we explain how the quark--photon interactions
are treated, leading to the correct results for the anomalous
processes. In the sixth section we present the results and
a discussion of them, whereas the seventh contains the 
concluding discussion on how the chosen model works.

\section{Bound-state models and axial anomaly}
\label{BoundAA}

It is not surprising that $q\bar q$ bound-state models have problems
in accounting for the processes dominated by the axial anomaly,
because their general philosophy is often such that the details
of the internal dynamics and structure of bound-states are given 
precedence over the correct incorporation of the symmetries of
the underlying theory which they would like to model. 
(The Nambu--Jona-Lasinio (NJL) model \cite{NJLa,NJLbc} applied to 
quarks -- reviewed by, {\it e.g.}, \cite{Sandy} -- 
is a notable exception, but due to its low momentum cutoff, 
its usual versions have other problems with 
the axial anomaly, which we will comment on later.) On the other
hand, the axial anomaly effects do not depend on the internal 
$q\bar q$ structure of pseudoscalar mesons at all, but appear
only if the light pseudoscalars are (pseudo)Goldstone bosons
of D$\chi$SB. Therefore, the constituent quark bound state approaches
which (unlike the consistently coupled SD-BS ones) do not incorporate
D$\chi$SB, cannot be consistent with the axial anomaly even if they 
successfully reproduce the magnitude of the $\pi^0\to\gamma\gamma$ 
width, because they introduce strong structure dependence where it 
should remain small. Namely, anomalous amplitude is independent of the
structure, while the non-anomalous effects on top of it should not exceed 
several percent \cite{Donoghue+alKnjiga} due to PCAC and 
Veltman--Sutherland theorem.
The incorporation of the axial anomaly in the fashion of, {\it e.g.}, 
Roberts \cite{Roberts}, Frank {\it et al.} \cite{Frank+al}, 
Burden {\it et al.} \cite{Burden+al96}, or Bando {\it et al.} \cite{bando94},
therefore represents a significant recent advance in the theory of bound 
states.

Approaches depicting the pion as a quark--antiquark bound state but 
not employing D$\chi$SB, do not succeed in capturing the effect of 
the axial anomaly. Namely, except through the fine-tuning of model 
parameters (as in Ref. \cite{Mu"nz+al94} for example), 
such modeling of the $q\bar q$ bound states has not provided 
even the numerical results for the $\pi^0\to\gamma\gamma$
decay width that can quite favorably match the experiment and
thus try to yield an explanation for the $\pi^0\to\gamma\gamma$
width that would be based on hadronic structure, as an alternative
to the physical understanding based on the Abelian axial anomaly.
For example, Horbatsch and Koniuk \cite{HK93} point out that
even with relativistic corrections, the constituent quark approaches
lead to $\pi^0\to\gamma\gamma$ decay widths exceeding the experimental  
decay width 
\begin{equation}
W_{exp}(\pi^0\to\gamma\gamma)=7.74\pm 0.56 \,\, {\rm eV}
\label{expPi0Width}
\end{equation}
by three orders of magnitude, unless {\it ad hoc} correction factors are 
introduced. Although Ackleh and Barnes \cite{AB92}, for example, use them 
for multiplying the $P\to\gamma\gamma$-widths, they themselves stress that 
these factors, being difficult to justify physically, are largely arbitrary 
and primarily motivated by comparison with experimental data.
On the other hand, the authors of Ref. \cite{HK93} reject this 
procedure, claiming that their improved approach gives
acceptable $\eta, \eta^\prime$ widths, the very large
discrepancy of 3 orders of magnitude still affecting only
the $\pi^0$-width. However, since the character of
$\gamma\gamma$-decays of these pseudoscalars
cannot be {\it drastically} different one from the other (as
all should be in varying degrees influenced by the axial anomaly),
their results on $\eta$-$\eta^\prime$
can be regarded only as parameter fittings without much
physical significance if the $\pi^0$-width is so very wrong
at the same time. 

Formulating calculations in the relativistically
covariant manner solves some of the problems from which 
not-fully-relativistic approaches were suffering. However,
improving the relativistic bound-state formalism in a
manner which leaves the treatment of $\pi^0\to\gamma\gamma$
structure- (model-)dependent also does not help decisively. 
{\it E.g.}, in the Bethe--Salpeter formalism without D$\chi$SB, 
Guiasu and Koniuk \cite{GuiasuKoniuk93} did obtain a 
suppression with respect to the 3-orders-of-magnitude overestimate 
of the $\pi^0\to\gamma\gamma$ width in the naive constituent model, 
but had to be satisfied with the correct order of magnitude, namely 
the width of roughly 4 eV; and the best (to our knowledge) result in 
those structure--dependent attempts was achieved by M\" unz {\it et al.} 
\cite{Mu"nz+al94}, who obtained -- in one of their fits -- the width of 
7.6 eV in a Bethe--Salpeter approach without D$\chi$SB, {\it i.e.}, 
with constituent quarks simply postulated, but at the expense
of fine--tuning the model parameters to very unlikely values which
may contradict with other empirical quantities. Most notably, they
are forced to suppose that the constituent $u(d)$-quark mass is quite 
light, $M_u = M_d = 170$ MeV, and such values of the {\it constituent}
quark mass are certainly unsatisfactory if tried in some other
applications, such as reproducing the spectrum of baryons.
Indeed, the later and broader fits to the meson spectrum in
practically the same approach \cite{Mu"nz96}, used more acceptable 
values of the parameters, but had to settle for
$W(\pi^0\to\gamma\gamma)$ between 3.81 eV and 5.07 eV. 
M\" unz {\it et al.} \cite{Mu"nz+al94} see their 
achievement in providing a quark bound-state alternative to 
the Goldstone-plus-anomaly picture (see their page 429). 
However, considering which alternative is more favorable (or rather, 
not ruled out) on phenomenological grounds, does not seem very 
essential. Namely, it is not just that D$\chi$SB in QCD is favored on 
phenomenological (and other) grounds -- it seems it is in fact 
obligatory, so there are probably no alternatives.
This is especially clearly stated by 'tHooft in Ref. \cite{'tHooft80}, 
according to which it is {\it not} possible that
in some variant of QCD, chiral symmetry is not spontaneously
(dynamically) broken. He stresses that in SU(N) binding theories,
chiral symmetry {\it must} be broken spontaneously, so that in the
chiral limit the anomaly must be reproduced by massless bound states
which are Goldstone modes of this spontaneous (dynamical) chiral symmetry 
breaking. Therefore, one should {\it reconcile} the bound-state picture
with the anomaly picture if one wants to have satisfactory understanding 
of the light pseudoscalar $q\bar q$ bound states and their interactions. 
Noting this need, already Hayne and Isgur \cite{HI82} 
expressed the hope (in their concluding section) that 
there may exist a massive-constituent model which is 
physically equivalent to the chiral-symmetry picture.

The woe affecting $q\bar q$ bound states when axial anomaly is at work,
makes one really appreciate the successes with anomalous processes
achieved by the methods incorporating the symmetries and anomalies
of QCD directly, like the chiral perturbation theory ($\chi$PT) 
with Wess--Zumino anomaly action. (See, {\it e.g.}, \cite{L95}.)
However, quarks are nowhere present explicitly in $\chi$PT, 
so that comparing constituent quark approaches with such a very
different, complementary approach, is not what is most frustrating.
The until recently unsatisfactory situation in the bound-state
approaches to $\pi^0\to \gamma\gamma$ was especially bothersome
if compared with the simple {\it free}-quark loop
(or even the old ``baryon loop") calculations of the triangle
diagram coupled to an ``elementary", structureless pion field,
which readily reproduced the anomaly amplitude $1/4\pi^2 f_\pi$
in the chiral limit {\it if} the Goldberger-Treiman relation is used.
(See, for example, \cite{itzykson80} or \cite{DeWittSmith}
for expositions of the connection between the simple 
free-quark loop calculation {\it a la} Steinberger and the 
$\sigma$-model anomaly analysis \cite{S49,Jackiw97}.)

Such calculations employing elementary pseudoscalar coupling to free 
quarks, thus seem to indicate that the bound--state
wave-functions that describe the internal $q\bar q$ sub-structure,
are a superfluous element that can only spoil reproducing the
axial anomaly effects when calculating the triangle diagram.
Does it, then, mean 
that the $\pi^0 \rightarrow \gamma\gamma$ decay
can somehow be the argument in favor of those (``the opponents
of the quark model", as Ref. \cite{HK93} puts it)
who reject the view of the pion as a $q\bar q$ bound state
in favor of the pion as a Goldstone boson of D$\chi$SB?
Certainly not, unless one insists that one must choose: {\it either} 
a quark model, {\it or} the Goldstone-plus-anomaly picture; 
our point is that in the literature
there already is at least one example \cite{jain91,munczek92,jain93b}
of a  constituent quark model which is very successful over a very
wide range of masses, and which can also solve the problems with 
anomaly because it incorporates the correct chiral symmetry
behavior through the coupled SD-BS mechanism:
the constituent quarks come about through
dynamical dressing in SD equations and the light $q\bar q$-solutions
of the BS equation in the {\it consistent approximation}
(here: rainbow-ladder) are (pseudo)Goldstone bosons!

As we will explain in the following sections, constituent models equivalent 
to the chiral symmetry picture are provided by the coupled SD-BS approach, 
because it simultaneously describes light pseudoscalar mesons as {\it both} 
bound states of constituent quarks {\it and} (pseudo)Goldstone bosons of 
D$\chi$SB, which produces these  constituent quarks in the first place. 
This gives the present treatment of $\gamma\gamma$-decays of pseudoscalar 
mesons a fundamental advantage over most of the other constituent quark 
approaches. For example, this enables us (essentially in the fashion of 
Bando {\it et al.} \cite{bando94} and Roberts and others 
\cite{Roberts,Frank+al} in the {\it Ansatz} approach)
to reproduce -- even analytically and exactly -- the famous, 
empirically successful result of the Abelian axial anomaly 
for the $\gamma\gamma$ width of $\pi^0$, the lightest pseudoscalar,
\begin{equation}
W(\pi^0 \rightarrow \gamma\gamma) =
      \frac{\alpha_{\rm em}^2}{64\pi^3} \, \frac{M_\pi^3}{f_\pi^2} \, ,
\label{AnomWidth}
\end{equation}
while the same approach can arbitrarily depart from the chiral limit
all the way to the heaviest mesons.

Obtaining this {\it form} (\ref{AnomWidth}), {\it i.e.}, this 
relationship between the width $W(\pi^0 \rightarrow \gamma\gamma)$
and the pion leptonic decay constant $f_\pi$, is possible thanks to 
the correct chiral symmetry behavior in the coupled SD-BS approach. 
However, this is not enough in the approaches which are
so predictive that, {\it e.g.}, the pion leptonic decay constant 
$f_\pi$ is also calculable. What is needed for obtaining the good 
value for the $\pi^0 \rightarrow \gamma\gamma$ width (\ref{AnomWidth})
consistently, is not just plugging in the empirical value of $f_\pi$,
but that the {\it predicted} $f_\pi$ is close to this empirical value. 
In the light of that, Jain and Munczek's model 
\cite{jain91,munczek92,jain93b} is especially favorable
variant of the coupled SD-BS approach, because it is simultaneously 
a very successful constituent quark model in the sense of reproducing
many meson masses from very light to very heavy ones, as well as the
leptonic decay constants of pseudoscalar mesons. For the concrete 
model and parameters \cite{jain93b} adopted in this work, we obtain 
$f_\pi=93.2$ MeV. Let us note that the pion decay constant is actually 
not very sensitive to parameter and model variations, as all Refs. 
\cite{jain91,munczek92,jain93b} obtain $f_\pi$ close to that value 
(and to the experimental one, $f_\pi^{exp}=92.4$ MeV) although they 
differ in details of modeling the infrared (IR) part of the gluon
propagator and in parameter values.

At this point we can sketch one of the reasons why the present 
approach is more suitable for treating the anomalous processes,
than the NJL approach is.
We make a more detailed comparison with a variant \cite{TO95}
of the NJL model elsewhere \cite{KlKe2}, but here we should 
address the most easily explainable reason,
because one might in principle remark that recent
coupled SD-BS approaches were not the first to fulfill
Hayne and Isgur's \cite{HI82} hope for existence of a $q\bar q$ model
equivalent to the chiral-symmetry picture, since already the 
NJL model has the correct chiral symmetry behavior and incorporates
the pion as the (pseudo-)Goldstone boson of D$\chi$SB. (In a sense, 
one can say that already the NJL model belongs to the class of the
coupled SD-BS approaches, but with a particularly schematic
interaction.) In particular, Eq. (\ref{AnomWidth}) seems --
on the first thought -- to be easily reproduced, because 
if in the chiral limit the Goldberger-Treiman relation at the 
quark level holds in the NJL model, the $\pi^0\to\gamma\gamma$ 
NJL-calculation ({\it e.g.}, in \cite{TO95}) effectively reduces 
to the simple free-quark loop calculation. The second thought, 
however, reveals the following problem: in contradistinction 
to Jain and Munczek's model, where the momentum cutoff is either 
not needed (in the chiral limit \cite{jain91}) or practically 
infinite \cite{munczek92,jain93b} in comparison with the 
relevant hadronic scales, the Nambu--Jona-Lasinio approach contains
a {\it low cutoff} ($\Lambda_{{\rm NJL}} < 1$ GeV). Nevertheless,  
its triangle diagram calculation of $\pi^0 \rightarrow \gamma\gamma$
of course reproduces the anomaly result (\ref{AnomWidth}) only if 
there is no cutoff!  As Langfeld {\it et al.} \cite{LKR96} pointed 
out, the low cutoff in the NJL model would affect
the anomalous $\pi^0\to\gamma\gamma$ width (\ref{AnomWidth}) 
by the factor $1/(1+M^2/\Lambda_{{\rm NJL}}^2)^2$. For the typical
value of the ratio of the constituent quark mass $M$ and the NJL-cutoff,
which is $M/\Lambda_{{\rm NJL}} \sim 0.5$, this implies the reduction
with respect to the empirically successful width (\ref{AnomWidth}) 
by 30\% to 40\%.

Although we perform a more concrete comparison with a specific NJL 
calculation \cite{TO95} in another paper \cite{KlKe2}, in this paper 
it will also -- in the following three sections -- gradually become clear
that the chosen bound-state approach \cite{jain91,munczek92,jain93b}
can be considered a NJL--inspired, but more sophisticated approach 
to mesons. Correspondingly, our approach to two-photon processes 
contains improvements (both in the conceptual consistency and in
the quantitative details) not only with respect to constituent 
models without D$\chi$SB, but also with respect to the NJL approach.

\section{The coupled Schwinger-Dyson and Bethe-Salpeter
approach, and Abelian anomaly}
\label{SDBSandAA}

\noindent

The coupled Schwinger-Dyson and Bethe-Salpeter (SD--BS) approach is 
the approach in which the Bethe--Salpeter (BS) equation employs the
quark propagator obtained by solving the Schwinger-Dyson (SD) equation,
and both equations employ the same interaction between quarks, normally 
defined by a (partially) modeled gluon propagator and the ladder 
approximation for the quark--gluon vertices. It is one of the most 
interesting applications of SD and BS equations to the physics of hadrons. 
Refs. \cite{RW,Miransky,PenningtonELFE,nuclth9807026} provide good reviews 
of the SD and BS equations in  the hadronic context -- including the 
coupled SD--BS approach, which started developing in the eighties
and the beginning of nineties; {\it e.g.}, see \cite{CahillRoberts,Cahill+al87,Praschifka+al87,Praschifka+al88,Praschifka+al89,StainsbyCahill92}
and other similar references that can be found in 
\cite{RW,Miransky,PenningtonELFE,nuclth9807026}.

So far, the most successful variant{\footnote{Recently, similarly successful 
descriptions of phenomenology have been achieved employing separable 
interaction \cite{CahillGunnerAdelaideWrkshp98,CahillprivateCom98}.}}
of the coupled SD--BS approach 
to the meson spectra and decay constants has been that of Jain and 
Munczek \cite{jain91,munczek92,jain93b}. Hence, we chose their model 
to use it in Ref. \cite{KeKl1} for the first time in the context of 
electromagnetic interactions (other than electromagnetic mass differences). 
While some aspects of the issues we discuss are model--dependent,
so that the presented results are specific to the Jain--Munczek model,
some are in fact common to all variants of the {\it consistently} 
coupled SD--BS approach, including some closely related approaches 
employing {\it Ans\" atze} for the dressed quark propagators.
In particular, this is so for anomalous processes of chiral pions,
as will become apparent below.

A crucial development for properly embedding the electromagnetic 
interactions in the context of the bound states composed of dynamically 
dressed quarks occurred when Bando {\it et al.} \cite{bando94} and 
Roberts \cite{Roberts} 
demonstrated how the Adler-Bell-Jackiw axial anomaly can be incorporated
in the framework of SD and BS equations, reproducing (in the chiral limit)
the famous anomaly result for $\pi^0\to\gamma\gamma$ analytically.
This was also extended \cite{Frank+al} to the off--shell case 
$\gamma^\star\pi^0\to\gamma$. However, these treatments are all
restricted to the chiral limit (and its immediate vicinity), {\it i.e.}, 
to quarks with vanishing (or almost vanishing) current masses $m$
and their pseudoscalar meson composites -- pions
with zero (or almost zero) mass $M_\pi$. 
In contradistinction to that, our application of Jain and Munczek's
model \cite{jain91,munczek92,jain93b} to electromagnetic processes,
is not subject to such limitations; while agreeing with 
\cite{Roberts,Frank+al,bando94} for $m\to 0$, it can also be 
applied for large quark masses $m$.

In the present paper we argue that in the light
of the latest experimental results on 
$\eta_c\to\gamma\gamma$ from CLEO \cite{CLEO1995}, elements of
these treatments \cite{bando94,Roberts,Frank+al} appear 
essential also for the understanding of the electromagnetic processes 
of mesons in a totally different regime, far away from the chiral limit.

Refs. \cite{Roberts,Frank+al} avoided solving the SD equation for 
the dressed quark propagator $S$ by using an {\it Ansatz} quark 
propagator. Then, thanks to working {\it in the chiral limit}
and {\it the soft limit} (where the momentum of the pion $p \to 0$),
they also automatically obtained the solution of the BS equation. 
Namely, in this limit,
when the chiral symmetry is not broken explicitly but spontaneously
(dynamically), and when pions must consequently appear as Goldstone bosons, 
the solution for the pion
bound-state vertex $\Gamma_\pi$, corresponding to the Goldstone pion,
is -- to order ${\cal O}(p^0)$ -- determined (see, {\it e.g.}, Ch. 9 in 
\cite{Miransky}) by the dressed quark propagator $S(q)$: 
	\begin{equation}
	S^{-1}(q)
	=
	A(q^2)\slashed{q} - B(q^2)~.
	\label{quark_propagator}
	\end{equation}
For the pion bound-state vertex $\Gamma_\pi$, Ref. \cite{Roberts,Frank+al} 
concretely used the solution (given in Eq. (\ref{ChLimSol}) immediately 
below) that is of zeroth order in the pion momentum $p$
(which is appropriate in the soft limit $p^\mu \to 0$).
It is interesting that such ${\cal{O}}(p^0)$ $\Gamma_\pi$ 
fully saturates the Adler-Bell-Jackiw axial anomaly \cite{bando94,Roberts}.
In the chiral limit, the pion decay constant $f_\pi$ gives \cite{JJ}
the normalization of $\Gamma_\pi$, whereas its ${\cal{O}}(p^0)$ 
piece (again: {\it in the chiral limit}) is proportional to $B(q^2)$ 
from Eq. (\ref{quark_propagator}):
        \begin{equation}
\Gamma_\pi(q;p^2\! =\! M_\pi^2\! =\! 0)\!
 =\! 2 \frac{\gamma_5 B(q^2)_{m=0}}{f_\pi} \, ,   \label{ChLimSol}
        \end{equation}
(where the flavor structure of pions has been suppressed).
Thanks to using the chiral-and-soft-limit solution (\ref{ChLimSol}) 
(and to satisfying the vector Ward-Takahashi identity of QED, on which 
point we elaborate later), Bando {\it et al.} \cite{bando94} and 
Roberts \cite{Roberts} {\it analytically} reproduced the famous 
axial-anomaly result {\it independently} of what precise {\it Ansatz}
has been used \cite{Roberts} for $S(q)$. 
Note that this implies the independence of this result also 
on the inter-quark interactions, determining both the quark
propagator $S(q)$ {\it and} the internal pion structure. Namely, 
even if one uses such an {\it Ansatz} for $S(q)$, one can in 
principle invert the SD equation (Eq. (\ref{SD-equation}) below)
for the quark propagator, and find the effective gluon propagator
$G^{\mu\nu}(k)$ that would lead to this $S(q)$. Since the reproduction 
of the axial-anomaly amplitude is independent of the {\it Ansatz}
for the quark propagator, it is consequently independent also of what 
interaction, defined by $G^{\mu\nu}(k)$, has formed the $\pi^0$ bound state. 

Whereas this {\it Ansatz} scheme works beautifully in the chiral limit 
and very close to it, one must obviously depart from it if one wants to 
consider heavier quarks. Already when strange ($s$) quarks are present, 
(\ref{ChLimSol}) can be regarded only as an ``exploratory {\em Ansatz}"  
\cite{Burden+al96}. For even heavier $c$-- and $b$--quarks, the whole 
concept of the chiral limit is of course useless even qualitatively.
The two--photon processes of neutral pseudoscalar mesons that are much
heavier than $\pi^0$ and cannot be described as Goldstone bosons, will 
depend strongly on their internal hadronic structure. Thus, one needs to 
solve the pertinent bound--state equation, which is  determined by the 
interaction between quarks. On the other hand, one should also have the 
axial anomaly incorporated correctly. This consistency requirement is 
satisfied by the coupled SD-BS approach developed in a series of papers 
by Jain and Munczek \cite{jain91,munczek92,jain93b}, because their 
treatment in the chiral limit yields pions as Goldstone bosons of dynamical 
chiral symmetry breaking (D$\chi$SB). On the other hand, their treatment 
has also reproduced \cite{jain91,munczek92,jain93b} almost the whole 
spectrum of meson masses, including those in the heavy-quark regime, and 
also the leptonic decay constants ($f_P$) of pseudoscalar mesons ($P$).
We are therefore motivated to use this so far successful approach for 
calculating other quantities.  In this paper we present the calculation 
of the $\pi^0, \eta_c$, and $\eta_b \rightarrow \gamma\gamma$ decay widths, 
as well as of the $\gamma^\star\pi^0$-to-$\gamma$ transition form factor.

\section{Obtaining $\lowercase{q}\bar{\lowercase{q}}$ bound-state solutions}
\label{QBARQ}

\subsection{Effective gluon propagator}
\noindent
We define the interaction kernel for the SD and BS equations
following Jain and Munczek \cite{jain91,munczek92,jain93b}:
{\it i.e.}, we use an effective, modeled Landau-gauge gluon 
propagator given by 

        \begin{equation}
  g_{\rm st}^{\, 2} \, C_F     G^{\mu\nu}(k) = G(-k^2)
        ( g^{\mu\nu} - \frac{k^\mu k^\nu}{k^2} )~,
	\label{gluon_propagator}
        \end{equation}
\noindent where we have indicated that our convention is such that
not only the strong coupling constant $g_{\rm st}$, but also $C_F$, 
the second Casimir invariant of the quark representation, are absorbed
into the function $G$. For the present case of SU(3)$_c$, where the 
group generators are $\lambda^a/2$, namely the (halved) Gell-Mann 
matrices, $C_F= \frac{4}{3}$.

It is essential that the effective propagator function $G$ is the 
sum of the perturbative contribution $G_{UV}$ and the nonperturbative 
contribution $G_{IR}$:

	\begin{equation}
	G(Q^2) = G_{UV}(Q^2) + G_{IR}(Q^2)~,\;\;(Q^2=-k^2)~.
\label{factorG}
        \end{equation}

The perturbative part $G_{UV}$ is required to reproduce correctly
the ultraviolet (UV) asymptotic behavior that unambiguously
follows from QCD in its high--energy, perturbative regime.
Therefore, this part must essentially be given 
-- up to the factor $1/Q^2$ -- 
by the running coupling constant $\alpha_{\rm st}(Q^2)$ which is 
well-known from perturbative QCD, so that $G_{UV}$ is in fact 
{\it not} modeled. 

{}From the renormalization group, in the spacelike region ($Q^2=-k^2$),

	\begin{equation}
	G_{UV}(Q^2)
	=
               4\pi C_F \frac{\alpha_{\rm st}(Q^2)}{Q^2}
	\approx
\frac{{ 4\pi^2 C_F } d}{Q^2 \ln(x_0+\frac{Q^2}{\Lambda_{QCD}^2})}
		\left\{
			1 +
                 b \, \frac{\ln[\ln(x_0+ \frac{Q^2}{\Lambda_{QCD}^2})]}
			       {\ln(x_0+ \frac{Q^2}{\Lambda_{QCD}^2})}
		\right\}~,
\label{gluon_UV}
	\end{equation}
where we employ the two--loop asymptotic expression for 
$\alpha_{\rm st}(Q^2)$, and where $d=12/(33-2N_f)$, 
$b=2\beta_2/\beta_1^2= 2(19N_f/12 -51/4)/(N_f/3 -11/2)^2$,  
and $N_f$ is the number of quark flavors. 
The parameter $x_0$ is the infrared cutoff, introduced to
regulate the logarithmic behavior of $G_{UV}$ as the 
values of $Q^2$ approach $\Lambda_{QCD}^2$,
the dimensional parameter of QCD. As in \cite{jain93b},
we use $x_0=10$, but this is not really important since the 
results are only very weakly sensitive to the values of $x_0$, 
as was already pointed out by \cite{jain93b}. Following \cite{jain93b}, 
we set $N_f=5$ and $\Lambda_{QCD}=228\,\mbox{\rm MeV}$.
Although the top quark has meanwhile been found, its 
mass scale is far above the range of momenta relevant 
for bound state calculations, and even above the value 
of the UV cutoff needed in the massive version of our 
SD equations (see below). Therefore, there is no 
need to revise $G_{UV}$ (\ref{gluon_UV}) to include $N_f=6$.
(On the other hand, choosing $N_f$ below 5 would not be
satisfactory because {\it (i)} the momentum range of the order 
of the $b$ quark mass still has non-negligible influence in our
bound-state calculations, {{\it (ii)} the $b$ quark mass is
below the UV cutoff used in our ``massive SD equations", and
{{\it (iii)} sometimes we need the solutions for relatively 
high momenta, {\it e.g.}, to be able to see the asymptotic 
behavior of the propagator functions $A(q^2)$ and $B(q^2)$ 
-- see Figs. 3, 4 and 5.)

$G_{UV}(Q^2)$ is depicted by the dashed line in Fig. 1.

The non-perturbative part $G_{IR}$ should describe the infrared
(IR) behavior. The infrared behavior of QCD is however still
very far from being understood, so this non-perturbative part
must be modeled. We choose $G_{IR}$ from Ref.~\cite{jain93b}:

	\begin{equation}
	G_{IR}(Q^2)
	=
                    { 4\pi^2 C_F }  \,a\,Q^2 \, \exp(-\mu Q^2),
	\label{gluon_IR}
	\end{equation}
\noindent  with their \cite{jain93b} parameters
$a=(0.387\,\GeV)^{-4}$ and $\mu=(0.510\,\GeV)^{-2}$.
This {\it Ansatz} for $G_{IR}$ is depicted by the dotted
line in Fig. 1. 

The whole effective propagator function $G$ that we use, is
depicted by the solid curve in Fig. 1 and Fig. 2.
It is interesting and indicative that the gluon 
propagator function $G(Q^2)$ (\ref{factorG})-(\ref{gluon_IR}) is 
reasonably similar{\footnote{We thank R. Cahill for pointing this 
out to us.}} to the effective gluon propagator functions obtained by 
Cahill and Gunner in \cite{CahillGunner95} (short--dashed line in 
Fig. 2) and \cite{CahillGunner97preprint} (dash--dotted line in Fig. 2) 
using a closely related approximation to QCD, called the Global Color Model 
\cite{CahillRoberts}, reviewed recently by Tandy \cite{Tandy97}. One 
observes a somewhat less close agreement with the effective gluon 
propagator (the long--dashed line in Fig. 2) employed by Savkli and 
Tabakin \cite{SavkliTabakin97} in another related approach, which, however, 
has not achieved such a broad fit to meson spectrum and decay constants as 
the approach of \cite{jain93b} with the propagator 
(\ref{gluon_propagator})--(\ref{gluon_IR}).
  
\subsection{Generating constituent quarks through SD equations}

In contradistinction to \cite{Roberts,Frank+al} for example,
we do not use any {\it Ans\" atze} for the functions $A(q^2)$ and 
$B(q^2)$ in the dressed quark propagators  (\ref{quark_propagator}). 
These propagators are obtained for various flavors by solving the SD 
equation in the ladder 
approximation ({\it i.e.}, with bare quark-gluon vertices):

        \begin{equation}
S^{-1}(p) = \slashed{p} -\widetilde{m} - i \,  g_{\rm st}^{\, 2}\, C_F 
\int \frac{d^4k}{(2\pi)^4}
                    \gamma^\mu S(k) \gamma^\nu G_{\mu\nu}(p-k)~,
       \label{SD-equation}
        \end{equation}

\noindent where $\widetilde{m}$ is the bare
mass term of the pertinent quark flavor,
breaking the chiral symmetry explicitly.

The case $\widetilde{m}=0$ corresponds to the chiral limit where
the current quark mass $m=0$, and where the constituent quark mass
$B(0)/A(0)$ stems exclusively from D$\chi$SB \cite{jain91}. 
(Of course, calling ``the constituent  mass" the value of the 
``momentum-dependent constituent  mass function" $B(q^2)/A(q^2)$ 
at exactly $q^2=0$ and not on some other low $q^2$, is a matter 
of a somewhat arbitrary choice. Another conventional choice 
({\it e.g.}, in \cite{MarisRoberts97PRC56,MarisRobertsNT9710062}) is 
to call the solution of $-q^2=B^2(q^2)/A^2(q^2)$ the Euclidean
constituent-quark 
mass squared. However, since this is just a matter of choosing one's 
terminology, we stick to that of Jain and Munczek \cite{munczek92}. 
 
With the assumption that $u$ and $d$ quarks are massless, which is an 
excellent approximation in the context of hadronic physics, solving of 
(\ref{SD-equation}) yields the solutions for $A(q^2)$ and $B(q^2)$, 
displayed in respective Fig. 3 and Fig. 4 by the solid lines.

In these figures we also compare them with $A(q^2)$ and $B(q^2)$ 
corresponding to the dressed propagator {\it Ans\" atze} of the 
references \cite{Roberts,Frank+al}, represented by the dashed lines. 
Our massless solutions lead to the constituent $u$ (and $d$) quark 
mass $B(0)/A(0)=356$ MeV. The ratio $B(q^2)/A(q^2)$, namely 
our momentum-dependent mass function $B(q^2)/A(q^2)$ is depicted 
in Fig. 5 by the solid line, and the dashed line represents the
analogous ratio formed from $A(q^2)$ and $B(q^2)$ corresponding 
to the {\it Ans\" atze} of Refs. \cite{Roberts,Frank+al}.

Obviously, our chiral--limit solutions for $A(q^2)$ and $B(q^2)$ 
differ a lot from the {\it Ans\" atze} of \cite{Roberts,Frank+al}. We
display and compare them for stressing one of the key points of the 
later sections, namely that {\it in the chiral limit}, what matters
is the correct implementation of D$\chi$SB and Ward identities, 
rather than the concrete forms of the propagator and bound state 
solutions. Of course, the situation changes as the quark masses 
grow appreciably, and approaches the simple constituent picture. 
Nevertheless, for the small quark masses appropriate for the 
realistically light pion, the physical situation is still dominantly 
determined by D$\chi$SB and the resulting (pseudo)Goldstone character 
of the pion -- and indeed, for the realistically massive $u$- and 
$d$-quarks and their lightest bound state $\pi$, the differences 
with respect to the chiral limit turn out to be small in the present 
approach.

When $\widetilde{m}\neq 0$, the SD equation (\ref{SD-equation})
must be regularized by a UV cutoff $\Lambda$ \cite{munczek92,jain93b},
and the bare mass $\widetilde{m}$ is in fact a cutoff--dependent quantity.
We adopted the parameters of \cite{jain93b}, where
(for $\Lambda=134$ GeV) $\widetilde{m}(\Lambda^2)$ is 
$3.1$ MeV for the isosymmetric $u$- and $d$-quarks,
$73$ MeV for $s$--quarks,
$680$ MeV for $c$--quarks, and $3.3$ GeV for $b$--quarks.
Solving of (\ref{SD-equation}) then yields the solutions
$A(q^2)$ and $B(q^2)$ for ``slightly massive" $u$- and $d$-quarks,
``intermediately massive" $s$-quarks, as well as
the solutions for the heavy quarks $c$ and $b$. 
We essentially reproduce the results of Ref. \cite{jain93b} 
(within the accuracy permitted by numerical uncertainties). The 
$A(q^2)$ and $B(q^2)$ solutions for $\widetilde{m}(\Lambda^2)\neq 0$ 
are displayed in Figs. 3 and 4 by dotted lines marked by $u,d$ and
$s,c$ and $b$, indicating which flavor a curve pertains to. For the 
lightest, $u$- and $d$-quarks (with $\widetilde{m}=3.1$ MeV), 
both $A(q^2)$ and $B(q^2)$ are only 
slightly above the curves representing our respective chiral-limit 
solutions. More precisely, the difference is then at most 1.4\% 
(at $q^2=0$) for $A(q^2)$, while for $B(q^2)$ the largest absolute 
value of the difference (again occurring at $q^2=0$) amounts to an excess
of 6.2\% over our chiral-limit solution. The excess quickly becomes 
much smaller above $-q^2=0.2$ GeV. Admittedly, at $-q^2$ above 2 GeV, 
the {\it relative} difference between the ``chiral" and 
``slightly massive" $B(q^2)$'s starts growing again because of the 
different asymptotic behaviors of these respective solutions. They 
are, respectively, $B(q^2)\sim [\ln(-q^2/\Lambda_{QCD}^2)]^{d-1}/q^2$ 
and $B(q^2) \sim 1/[\ln(-q^2/\Lambda_{QCD}^2)]^d$, and are consistent
with the asymptotic freedom of QCD \cite{Lane74,Politzer76}. 
(This in turn results 
in the asymptotic behavior of the momentum-dependent, dynamical mass 
functions $B(q^2)/A(q^2)$, which is in accord with the behavior in 
perturbative QCD \cite{jain91,munczek92,jain93b,Politzer76,Tarrach81}). 
However, the absolute values of these $B(q^2)$'s (even for the 
``slightly massive" case) and of their difference are already very 
small at $-q^2>2$ GeV. 

For the more massive flavors, the deep Euclidean asymptotic behavior 
$B(q^2) \sim 1/[\ln(-q^2/\Lambda_{QCD}^2)]^d$ is also fulfilled, but of
course with very different coefficients (which are essentially 
proportional to the current quark masses \cite{munczek92,jain93b}). 
Also, $A(q^2)\to 1$ for all flavors as $-q^2\to\infty$. For
low $-q^2$, however, $A(q^2)$'s belonging to different flavors
exhibit interesting differences. The bump that characterizes 
the least massive (or chiral) $u,d$-quarks is absent already 
in $A(q^2)$ of our ``intermediately massive" $s$-quark, for which 
the fall-off is almost monotonical, as the increase (around $-q^2\sim 0.1$ 
GeV) above the $A(0)$-value is practically imperceptibly small. 
Moreover, for even heavier $c-$ and especially $b-$quarks, the
$A(q^2)$-values for even lowest $-q^2$'s, are below the corresponding
values of the chiral-limit $A(q^2)$. Comparing the various 
$A(q^2)$- and $B(q^2)$-solutions illustrates well how the 
importance of the dynamical dressing decreases as one considers
increasingly massive quark flavors. 

These $\widetilde{m}\neq 0$ solutions give us the constituent mass 
$B(0)/A(0)$ of 375 MeV for the (isosymmetric) $u$- and $d$-quarks, 
610 MeV for the $s$-quarks,
$1.54$ GeV for the $c$--quarks, and $4.77$ GeV for the $b$--quarks.
These are very reasonable values. Also, the momentum-dependent 
mass functions $B(q^2)/A(q^2)$ -- depicted in Fig. 5 -- in the 
presently chosen variant 
of the coupled SD-BS approach \cite{jain91,munczek92,jain93b} 
behave for all flavors in the way which correctly captures 
the differences between heavy and light quarks (stressed recently 
by Ref. \cite{MarisRobertsNT9710062}).

\subsection{Bound states of dynamically dressed constituent quarks}

In the chiral limit, 
Eqs. (\ref{quark_propagator})--(\ref{ChLimSol}) reflect the fact that 
solving of (\ref{SD-equation}) with $\widetilde{m}=0$ is already 
sufficient to give us the Goldstone pion bound-state vertex
(to ${\cal O}(p^0)$) which saturates the anomalous 
$\pi^0\to\gamma\gamma$ decay \cite{Roberts,bando94}. Of course, the 
usefulness of avoiding to solve the BS equation this way, diminishes 
with growing masses, and for heavier $q\bar q$ composites, such as
presently interesting $\eta_c$ and $\eta_b$, this does not make sense 
even qualitatively. In addition, in this paper we are interested in 
the effects -- however small -- of the finite masses of the light 
quarks and of the pion. Therefore, the massive pseudoscalar ($P$)
bound-state vertices $\Gamma_{P}$ for the massive pion, $\eta_c$, 
$\eta_b$, and also for the unphysical pseudoscalar $s\bar s$ bound 
state  $\eta_s$, will be obtained
in the same way as the bound-state vertex $\Gamma_{\cal M}$ 
of any other meson ${\cal M}$, by solving explicitly
the homogeneous BS equation

        \begin{equation}
                \Gamma_{\cal M}(q,p) 
                = i \,  g_{\rm st}^{\, 2} \, C_F 
                \int \frac{d^4q^\prime}{(2\pi)^4}
                \gamma^\mu S(q^\prime + \frac{p}{2})
      \Gamma_{\cal M}(q^\prime,p)   S(q^\prime - \frac{p}{2})
                \gamma^\nu
                G_{\mu\nu}(q-q^\prime)~.
        \label{BSE}
        \end{equation}

\noindent Here we have written the BS equation again in the 
ladder approximation, consistently with (\ref{SD-equation}). 
Note that $S$ is the quark propagator obtained by solving the 
SD equation (\ref{SD-equation}) with the same gluon propagator 
$G_{\mu\nu}$. 

For pseudoscalar mesons (${\cal M}=P$),
the complete decomposition of the BS bound-state vertex $\Gamma_P$
in terms of the scalar functions $\Gamma^P_i$ is:

        \begin{equation}
\Gamma_P(q,p)=\gamma_5 \left\{\, \Gamma^P_0(q,p) + \slashp \, \Gamma^P_1(q,p)
  + \slashq \, \Gamma^P_2(q,p) + [\slashp,\slashq]\, \Gamma^P_3(q,p)\, \right\}.
\label{decompGamma}
        \end{equation}  

\noindent (The flavor indices are again suppressed.)
The BS equation (\ref{BSE}) leads
to a coupled set of integral equations for the functions
$\Gamma^P_i$ $(i=0,...,3)$,
which can most easily be solved numerically in the Euclidean space. 

The formulations of the relativistic bound--state problem
through the ``amputated" BS vertices $\Gamma_{\cal M}(q,p)$
or the ``unamputated" BS {\it amplitudes} $\chi_{\cal M}(q,p)$,
mutually related through 
\begin{equation}
  \Gamma_{\cal M}(q,p) = S^{-1}(q+\frac{p}{2}) \chi_{\cal M}(q,p)
S^{-1}(q-\frac{p}{2})~,
        \label{BS-amplitude-amp.}
        \end{equation}
are of course equivalent, but we find that it is technically somewhat 
more convenient to solve the BS equation for the BS amplitude, 
        \begin{equation}
                S^{-1}(q+\frac{p}{2})
                \chi_{\cal M}(q,p)
                S^{-1}(q-\frac{p}{2})
                =
                i \,  g_{\rm st}^{\, 2} \, C_F \,
                \int \frac{d^4q^\prime}{(2\pi)^4}
                \gamma^\mu
                \chi_{\cal M}(q^\prime,p)
                \gamma^\nu
                G_{\mu\nu}(q-q^\prime)~,
        \label{amplitBSE}
        \end{equation}
as in \cite{jain91,munczek92,jain93b}.
For pseudoscalar 
mesons, the decomposition of the BS amplitudes in terms of the 
scalar functions $\chi^P_i$ proceeds in the same way as in 
(\ref{decompGamma}) for the pseudoscalar bound-state vertices:
        \begin{equation}
\chi_P(q,p)=\gamma_5 \left\{\, \chi^P_0(q,p) + \slashp \,
\chi^P_1(q,p)
  + \slashq \, \chi^P_2(q,p) + [\slashp,\slashq]\, \chi^P_3(q,p)\,
\right\}.
\label{decomp-chi}
        \end{equation}

Substitution of (\ref{decomp-chi}) into (\ref{amplitBSE}) results 
in a coupled set of integral equations for the functions $\chi^P_i$
($i=0,...,3$), which we solve in the Euclidean space for the massive
pion, $\eta_c$ and $\eta_b$. We also solve them for the 
$s\bar s$-pseudoscalar ($\eta_s$) even though it is unphysical, because it 
is needed for constructing $\eta$ and $\eta^\prime$ in Ref. \cite{KlKe2}.
($\eta$ and $\eta^\prime$ and their two-photon interactions are 
of similar interest to us as those of $\eta_c$ and $\eta_b$. However, 
obtaining the bound--state description for $\eta$ and $\eta^\prime$ 
is much more complicated owing to their mixing, so we treat them 
separately in another paper \cite{KlKe2}.)

In our calculations -- following Jain and Munczek 
\cite{jain91,munczek92,jain93b} --
we use the Chebyshev-polynomial-decomposition of the scalar functions 
$\chi^P_i$ ($i=0,...,3$) appearing in Eq. (\ref{decomp-chi}). This way 
one avoids the angular integration not only in the integral equations 
for the functions $\chi^P_i$ themselves, but also in applications of
these solutions, such as in our evaluation of the amplitudes for the 
$P\gamma\gamma$-processes addressed in the next section. 
While using the kernel and parameters of the Ref. \cite{jain93b} 
which often kept only the zeroth order Chebyshev moment 
because it was adequate for the spectrum of the pseudoscalars, we always
retain the first Chebyshev moment too. The accuracy of our procedure
has recently received an independent confirmation from 
Maris and Roberts \cite{MarisRoberts97PRC56}. In their study of
$\pi$- and $K$-meson BS amplitudes (or, quite precisely, the BS-vertices
(\ref{decompGamma})), they employed both the Chebyshev decomposition, and 
straightforward multidimensional integration.
Comparison of these two techniques confirms the very quick convergence
of the Chebyshev expansion \cite{MarisRoberts97PRC56}: in the case of 
equal quark and antiquark masses, such as in the pion, the zeroth and 
the first Chebyshev moment are enough for an accurate representation of 
the solution, while for the kaon, in spite of the difference in the masses 
of the constituents, just one more is needed. 

We confirm the success in reproduction of many
meson masses, especially of heavy ones, of the Jain--Munczek
approach \cite{jain91,munczek92,jain93b}, some small differences 
being due to our independent numerical procedures. Solving for 
$\chi_{P}$ ($ P = \pi , \eta_s , \eta_c , \eta_b $) using the 
parameter values fixed in Ref. \cite{jain93b}, we obtain the following. 
{\it i)} $M_{\pi} = 140$ MeV -- {\it i.e.}, less than 2\% away from the 
average over the empirical masses of the three pion isospin states, which 
value (138 MeV) is the most appropriate pion mass in the isospin limit. 
{\it ii)} $M_{\eta_s} = 721$ MeV, which value contributes to reasonable 
predictions for the  masses of physical states $\eta$ and $\eta^\prime$ in Ref. 
\cite{KlKe2}. {\it iii)} Solving for $\chi_{\eta_c}$, we get 
$M_{\eta_c}=2.875$ GeV, whereas the experimental $\eta_c$-mass 
is $M_{\eta_c}^{exp}=2.979$ GeV.   {\it iv) } For $\eta_b$, where 
there are no experimental results yet, we predict $M_{\eta_b}=9.463$ GeV.

The amputated BS bound-state vertices $\Gamma_{P}$
are sometimes technically more convenient for describing 
some aspects of the presently interesting processes 
involving bound states, than the equivalent formulation 
through the BS amplitude $\chi_{P}$.
The derivation of the closed-form expression (Eq. (\ref{AnomAmp}) 
below) for the amplitude of the ABJ anomaly--induced decay 
$\pi^0\to\gamma\gamma$, is in the chiral limit therefore performed 
this way, employing the amputated vertex $\Gamma_\pi$.
The reason why this is simpler is clear if we recall that 
\cite{bando94,Roberts} proved that the ${\cal{O}}(p^0)$-piece 
of the amputated pion bound-state vertex, namely $\Gamma^{\pi}_0(q,0)$
in (\ref{decompGamma}), given in the chiral limit by (\ref{ChLimSol}), 
fully saturates the axial anomaly. 
In contrast to that, using the BS-amplitude here would introduce 
unnecessary complications since the connection between the functions 
$\Gamma^{\pi}_i$ and $\chi^{\pi}_i$ (defined by (\ref{BS-amplitude-amp.}))
is such that $\Gamma^{\pi}_0(q,0)$ is tied to all four functions
$\chi^{\pi}_i(q,0)$. On the other hand, we find the {\it un}amputated
BS amplitudes $\chi_{\cal M}$ generally more suitable for solving 
BS equations on the computer, since the number of quark propagator lines
is reduced in the integral
-- see the BS equations (\ref{BSE}) and (\ref{amplitBSE}).

\section{
$\pi^0, \eta_{\lowercase{s}}, \eta_{\lowercase{c}}, 
\eta_{\lowercase{b}} \rightarrow \gamma\gamma$,
and $\gamma^\star\pi^0 \rightarrow \gamma$ processes}

\noindent

The transition matrix element for all of the processes
$P \rightarrow \gamma\gamma$ ($P = \pi^0, \eta_s, \eta_c, \eta_b$) 
and $\gamma^\star\pi^0 \rightarrow \gamma$ has the form,

        \begin{equation}
        S_{fi}
        =
        (2\pi)^4 \delta^{(4)}(p-k-k^\prime)
        \varepsilon^{\mu\star}(k,\lambda)
        \varepsilon^{\nu\star}(k^\prime,\lambda^\prime)
        ie^2 \,
        T_P^{\mu\nu}(k,k^\prime)~,
        \label{sfi2}
        \end{equation}

        \noindent where

        \begin{equation}
        T_P^{\mu\nu}(k,k^\prime)
        =
        i\int d^4x\, \exp(ik\!\cdot\! x)
        \langle 0|T \{ J^\mu(x) J^\nu(0) \} |P(p)\rangle~,
        \label{Tmunu}
        \end{equation}

\noindent and where $J^\mu(x)$ is the electromagnetic current of quarks,

        \begin{equation}
        J^\mu(x)
        =
        \bar\psi(x) \gamma^\mu {\cal Q} \psi(x)~,
        \end{equation}

\noindent and ${\cal Q} = 
{\rm diag}({\cal Q}_u, {\cal Q}_d, {\cal Q}_s, {\cal Q}_c, {\cal Q}_b) =
{\rm diag}(+\frac{2}{3},-\frac{1}{3},-\frac{1}{3},+\frac{2}{3},-\frac{1}{3})$ 
is the quark charge operator in units of the proton charge $e$.
The photon momenta are $k$ and $k^\prime$, while $p=k+k^\prime$ is
the meson momentum.

For $\pi^0, \eta_s, \eta_c$ and $\eta_b\to\gamma\gamma$, the both 
photons ($\gamma$) are real, with $k^2=k^{\prime 2}=0$, and consequently
$\varepsilon^{\mu}(k,\lambda)$ and
$\varepsilon^{\nu}(k^\prime,\lambda^\prime)$
are both just the photon polarization vectors. In contrast,
for $\gamma^\star\pi^0\to\gamma$, only one photon is real ($k^2=0$),
whereas the other is virtual, stemming for example from the electron
current like in TJNAF. Only $\varepsilon^{\mu}(k,\lambda)$ is 
the photon polarization vector, whereas
$\varepsilon^{\nu}(k^\prime,\lambda^\prime)$
should be understood as the product of the electron electromagnetic
current
and the photon propagator transferring the momentum $k^\prime$,
$k^{\prime 2}=-Q^2\neq 0$. In any case, the problem boils down to
computing the
tensor $T^{\mu\nu}(k,k^\prime)$ defined by Eq.~(\ref{Tmunu}), {\it
i.e.},
the $P$--to--vacuum matrix element of electromagnetic quark currents.
By symmetry arguments, $T_P^{\mu\nu}$ can be written as \cite{itzykson80}

        \begin{equation}
        T_P^{\mu\nu}(k,k^\prime)
        =
        \varepsilon^{\alpha\beta\mu\nu} k_\alpha k^\prime_\beta
        T_P(k^2,k^{\prime 2})~,
        \label{Tmunu-parametrization}
        \end{equation}

\noindent where $T_P(k^2,k^{\prime 2})$ is a scalar, a function of the
scalar
products of the momenta. For $\pi^0\to\gamma\gamma$ as well as
$\eta_c\to\gamma\gamma$ and $\eta_b\to\gamma\gamma$, where all particles
are real, {\it i.e.}, on their mass shells, we have simply
$T_P(0,0)=\mbox{\rm const}$ in (\ref{Tmunu-parametrization}).
After summing over polarizations and integration over the photon phase
space we finally get the width

        \begin{equation}
        W(P\to\gamma\gamma)
        =
        \frac{\pi\alpha_{\rm em}^2 M_P^3}{4}
     \, |T_P(0,0)|^2~, \qquad (P = \pi^0, \eta_c, \eta_b).
        \label{PiWidths}
        \end{equation}

\noindent 
On the other hand, for $\gamma^\star\pi^0\to\gamma$ transition,
only the final photon is necessarily on the mass shell, $k^2=0$,
whereas $k^{\prime 2}=-Q^2$ is the square of the momentum
transferred from the projectile
beam. In general, the pion can be off--mass shell, with $p^2\neq
M_\pi^2$. Ref. \cite{Frank+al} discusses how to handle also this 
more general possibility. In the present paper, we limit ourselves to 
the $\gamma^\star\pi^0\to\gamma$ transition form factor $F(Q^2)$ on
the $\pi^0$ mass shell, defined by

        \begin{equation}
        F(Q^2)
        =
        \frac{T_{\pi^0}(0,-Q^2)}{T_{\pi^0}(0,0)}~.
\label{transFF}
        \end{equation}

How do we go about calculating $T_P$, or, equivalently, $T_P^{\mu\nu}$?
We adopt the framework advocated by (for example) 
\cite{bando94,Roberts,Frank+al,Burden+al96}
in the context of electromagnetic interactions of BS bound states,
and called the generalized impulse approximation (GIA) by 
\cite{Roberts,Frank+al,Burden+al96}.
This means that in the triangle graph we use the
{\it dressed} quark propagator $S(q)$, Eq.~(\ref{quark_propagator}),
and the pseudoscalar BS bound--state vertex $\Gamma_P(q,p)$
instead of the bare $\gamma_5$ vertex. Another essential 
element of the GIA is to use an appropriately dressed {\it electromagnetic}
vertex $\Gamma^\mu(q^\prime,q)$, which satisfies the vector Ward--Takahashi
identity,

        \begin{equation}
        (q^\prime-q)_\mu \Gamma^\mu(q^\prime,q) =
                S^{-1}(q^\prime) - S^{-1}(q)~.
        \label{WTI-v}
        \end{equation}

\noindent Assuming that photons
couple to quarks through the bare vertex $\gamma^\mu$
would be inconsistent with
our quark propagator, which, dynamically dressed through
Eq.~(\ref{SD-equation}),
contains the momentum-dependent functions $A(q^2)$ and $B(q^2)$.
The bare vertex $\gamma^\mu$ obviously violates (\ref{WTI-v}),
implying the non-conservation of the electromagnetic vector current
and of the electric charge. Since no-one has yet satisfactorily 
solved the pertinent SD equation for the 
dressed quark-photon vertex $\Gamma^\mu$, it is
customary to use realistic {\it Ans\"{a}tze}, in the development
of which a number of researchers invested much effort.
Motivated by its earlier successful usage in some related work,
such as \cite{Roberts} or \cite{Frank+al}, we choose the Ball--Chiu 
(BC) ~\cite{BC} vertex for $\Gamma^\mu(q^\prime,q)$:

        \begin{eqnarray}
        \Gamma^\mu(q^\prime,q) =
        A_{\bf +}(q^{\prime 2},q^2)
       \frac{\gamma^\mu}{\textstyle 2}
        + \frac{\textstyle (q^\prime+q)^\mu }
               {\textstyle (q^{\prime 2} - q^2) }
        \{A_{\bf -}(q^{\prime 2},q^2)
        \frac{\textstyle (\slashed{q}^\prime + \slashed{q}) }{\textstyle 2}
         - B_{\bf -}(q^{\prime 2},q^2) \}~,
        \label{BC-vertex}
        \end{eqnarray}

\noindent where
$H_{\bf \pm}(q^{\prime 2},q^2)\equiv [ H(q^{\prime 2}) \pm H(q^2) ]$,
for $H = A$ or $B$. 
Obviously, this {\it Ansatz} does not introduce any
new parameters as it is completely determined by the quark
propagator (\ref{quark_propagator}). Its four chief virtues, however,
are {\it (i)} that it satisfies the Ward--Takahashi identity (\ref{WTI-v}),
{\it (ii)} that it reduces to the bare vertex in the free-field limit
as must be in perturbation theory, {\it (iii)} that its transformation
properties under
Lorentz transformations and charge conjugation are the same as for the
bare vertex, and {\it (iv)} it has no kinematic singularities.

It is important to note that the correct
axial-anomaly result cannot be obtained \cite{Roberts}
analytically in the chiral limit unless 
a quark--photon--quark ($qq\gamma$) vertex
that satisfies the Ward-Takahashi identity 
is used (even if D$\chi$SB {\it is} employed and the
pion {\it does} appear as a Goldstone boson, as
comparison with \cite{horvat91,horvat95} shows).

In the case of $\pi^0$ for example, the GIA yields 
the amplitude $T_P^{\mu\nu}(k,k^\prime)$:

\begin{displaymath}
        T_{\pi^0}^{\mu\nu}(k,k^\prime)
        =
        -
        N_c \, 
        \frac{{\cal Q}_u^2 - {\cal Q}_d^2}{2}
        \int\frac{d^4q}{(2\pi)^4} \mbox{\rm tr} \{
        \Gamma^\mu(q-\frac{p}{2},k+q-\frac{p}{2})
        S(k+q-\frac{p}{2})
\end{displaymath}
\begin{equation}
	  \qquad
        \times
        \Gamma^\nu(k+q-\frac{p}{2},q+\frac{p}{2})
        S(q+\frac{p}{2})
        \Gamma_{\pi^0}(q,p)
        S(q-\frac{p}{2}) \}
        +
        (k\leftrightarrow k^\prime,\mu\leftrightarrow\nu).
\label{Tmunu(2)}
\end{equation}

\noindent (The analogous expressions for 
$\eta_c$ and $\eta_b$, or any other neutral pseudoscalar,
are straightforward.)
The number of colors $N_c$ arose from the trace over the color 
indices. The $u$ and $d$ quark charges in units of $e$,
${\cal Q}_u=2/3$ and ${\cal Q}_d=-1/3$, appeared from tracing the product
of the the quark charge operators ${\cal Q}$ and $\tau_3/2$,
the flavor matrix appropriate for $\pi^0$:
${\rm tr}({\cal Q}^2\tau_3/2)= ({\cal Q}_u^2 - {\cal Q}_d^2)/2$.

\section{Discussion of the results}

\noindent
In the chiral and soft limits, the present approach yields the 
{\it form} of the amplitude for $\pi^0\to\gamma\gamma$ which is 
completely independent of the pion structure. Namely, our framework 
is in this limit equivalent to \cite{bando94,Roberts,Frank+al}, as 
demonstrated by the fact that, in the chiral limit, 
with the ${\cal O}(p^0)$ solution (\ref{ChLimSol})
but independently of its concrete shape $B(q^2)$, we too can 
reproduce analytically, in the closed form, the famous ``triangle"
anomaly amplitude{\footnote{By the way, our approach in the chiral  
and soft limit also reproduces analytically the ``box" anomaly 
result for $\gamma^\star\to\pi\pi\pi$, in the fashion of Alkofer
and Roberts \cite{AR96} in the {\it Ansatz} approach.}} 
\begin{equation}
T^{chiral}_{\pi^0}(0,0) \, = \, \frac{1}{4\pi^2 f_\pi} \, .  
\label{AnomAmp}
\end{equation}
The decay width (\ref{PiWidths}) is then given by Eq. (\ref{AnomWidth}), 
and its numerical value is in excellent agreement with experiment (see 
Table~1), {\it if} the pion decay constant $f_\pi$ is also predicted 
correctly. 

As already pointed out above, this result is independent
of our concrete choice of the interaction kernel and the
resulting hadronic structure of $\pi^0$; this is as it should be,
because the axial anomaly, which dominates $\pi^0\to\gamma\gamma$,
is of course independent of the structure.

It is then not surprising that the calculations
for $\pi^0\to\gamma\gamma$ which rely on the details
of the hadronic structure (be it in the context of the BS
equation, nonrelativistic quarks, or otherwise) fail
to describe this decay accurately even when the model 
parameters are fine-tuned for that purpose.

On the other hand, we should point out that Jain and Munczek's
\cite{jain91,munczek92,jain93b} approach to the bound states
describes the processes such as the anomaly--dominated 
$\pi^0\to\gamma\gamma$ or $\gamma^\star\to\pi\pi\pi$, 
more consistently than some other approaches 
which also analytically obtain the correct {\it expressions} 
(\ref{AnomAmp}) and (\ref{AnomWidth}). The reason is 
that their \cite{jain91,munczek92,jain93b} treatment 
of the all--important pion decay constant is exceptionally 
consistent. Namely, in the chiral limit, the axial-vector 
Ward--Takahashi identity  
         \begin{equation}
        (q^\prime-q)_\mu \Gamma^{a\mu}_5(q^\prime,q) =
        \frac{1}{2}\lambda^a 
        [S^{-1}(q^\prime)\gamma_5 + \gamma_5 S^{-1}(q)]
        \label{WTI-a}
        \end{equation}
requires that the normalization{\footnote{As in \cite{KlKe2}, the 
normalization is in this paper defined as in Eq. (2.8) of Ref. 
\cite{munczek92}, up to the factor of $N_c$, since we do not adopt 
their \cite{munczek92} conventional color factor of $1/\sqrt{N_c}$ 
in the definition of BS amplitudes.}}
${\cal N}_\pi$ of the BS pion bound state must be equal to the pion 
decay constant \cite{JJ}, ${\cal N}_\pi = f_\pi$.
However, as pointed out recently by Maris, Roberts, and Tandy
\cite{MarisRoberts97PRC56,MarisRobertsTandy97}, this equality is 
{\it not} obeyed in model studies to date (unless one unrealistically
assumes $A(q^2)\equiv 1$) because of neglecting the ${\cal O}(p)$
terms in the decomposition of the BS amplitudes (\ref{decomp-chi}) 
or bound-state vertices (\ref{decompGamma}).
Nevertheless, the approach of Jain and Munczek is a notable
exception, since it {\it does}
satisfy this important consistency relation very accurately,
within 2\% to 3\%. This was pointed out already in their first
paper \cite{jain91} devoted precisely to the consistent calculation 
of $f_\pi$ in the SD--BS framework, where they stressed the 
importance of subleading Dirac components. 
This fact seems to have been largely overlooked, and we point
it out especially because of its obvious favorable implications 
for the consistent description of $\gamma\gamma$-processes 
performed here.

The off-shell extension of $\pi^0\to\gamma\gamma$, namely 
$\gamma^\star\pi^0\to\gamma$, is described by the transition form 
factor $F(Q^2)$ (\ref{transFF}) which we calculated for space--like 
($Q^2 > 0$) transferred momenta. As in \cite{Frank+al}, we have used
the chiral-limit solution (\ref{ChLimSol}) for the pion bound state, but
we have obtained $B(q^2)$ by solving the SD equation (\ref{SD-equation})
with a
specified kernel, concretely (\ref{gluon_propagator})--(\ref{gluon_IR}).
We compare our $F(Q^2)$ in Fig.~7 with the CELLO data \cite{behrend91}
and with the $F(Q^2)$ of \cite{Frank+al}, which is a 
quark-propagator-{\it Ansatz} approach but related
to ours. We also display the Brodsky--Lepage interpolation
\cite{BrodskyLepage} to the perturbative QCD factorization limit
$Q^2 F(Q^2)\rightarrow 8\pi^2 f_\pi = 0.67$ GeV$^2$, and the 
curve resulting from the vector-meson-dominance type of approach 
of \cite{BH}.
Although our solution for $B(q^2)$, parameterizing the pion 
structure in the chiral limit, is significantly different
from the {\it Ansatz} used by \cite{Frank+al} (see Fig.~4),
our curve for $F(Q^2)$ is only marginally better than that of
\cite{Frank+al}, and we conclude that for $Q^2 < 2.5$ GeV$^2$ 
these curves fit the CELLO experimental points{\footnote{Our
comparison with the new CLEO data \cite{gronberg98} at higher
momenta up to $Q^2 \sim 8$ GeV$^2$, as well as the asymptotic
behavior of the $\gamma^\star\pi^0\to\gamma$ form factor, 
is in preparation \cite{ZTF98-04}.}
with comparable quality.
The ``interaction size'', defined by \cite{Frank+al} via
$\langle r^2_{\gamma\pi^0\gamma} \rangle = - 6 F^\prime(Q^2)_{Q^2=0}$,
discriminates even less between our respective $B(q^2)$'s;
namely, our result of 0.46 fm is practically the same as
${\langle r^2_{\gamma\pi^0\gamma} \rangle}^{1/2}=0.47$ fm
of \cite{Frank+al}. (The monopole fit to the CELLO data
\cite{behrend91} yields $0.65 \pm 0.03$ fm.)
It seems, therefore, that the differences in the pion structure
play a relatively small role for the $\gamma^\star\pi^0\to\gamma$
transition form factor, at least in the chiral limit, although 
cannot be reduced to the dependence of $f_\pi$ as for the 
on-shell $\gamma\gamma$ amplitude of the (pseudo)Goldstone $\pi^0$.

Of course, the situation is very different for quark masses
of the order of $\Lambda_{QCD}$ and higher: only the numerical
evaluation of $T_P(0,0)$ is reliable in this regime.  
Moreover, the details of the chosen interaction kernel
and the resulting propagator functions $A(q^2)$ and $B(q^2)$,
as well as the bound state solutions, {\it do} matter in that regime
(which gets further and further away from the domination of the 
axial anomaly with growing quark masses, as illustrated
by the amplitude ratios in the last column of Table~1).
This is naturally the case with the heavy-quark composites 
$\eta_c$ and $\eta_b$. Their two--photon widths are also given 
in Table~1. In the case of $\eta_b$, 
only predictions exist, as there are no experimental data yet,
so that the calculated mass of $\eta_b$ had to be used in the 
phase-space factors, unlike $\pi^0$ and $\eta_c$. 

The case of $\eta_c$ is very intriguing. The widths
$W(\eta_c\to\gamma\gamma)$ resulting from
the nonrelativistic constituent potential models had
long seemed to be in good agreement with experiment, but 
have more recently been shown \cite{AhmadyMendel} to
rise to far too large values of $11.8\pm 0.8\pm 0.6$ keV
after the calculations have been {\it improved} by removing
certain approximations.
The estimates \cite{AhmadyMendel} of the relativistic corrections
indicate that they are so large that they can reduce the width 
back down to around 8.8 keV. Such strong relativistic corrections 
corroborate the view that the relativistic approaches to bound 
states retain their importance on the mass scale of the 
$c$--quarks. Since the latest (1996) \cite{PDG96} Particle 
Data Group (PDG) average has also risen with respect to the 
earlier (1994) PDG average \cite{PDG94} up to 
$W^{PDG}_{'96}(\eta_c\to\gamma\gamma)= 7.5\pm 1.6$ keV,
it can appear that the substantial relativistic corrections
have indeed managed to solve the problem, bringing about the 
agreement with experiment. However, if this is so, then {\it 
fully relativistic} treatments through the BS equation should 
be able to yield results which are at least as good --
{\it i.e.}, in no worse agreement with the experimental
$\eta_c\to\gamma\gamma$ width than the relativistically
corrected width from the nonrelativistic constituent potential 
models \cite{AhmadyMendel}. Only if this is the case, we can 
say we understand $\eta_c\to \gamma\gamma$ simply through 
the relativistic effects in constituent models.

Nevertheless, the situation is not like that. 
If one surveys the theoretical treatments
of the two--photon physics of mesonic $q\bar q$--composites
attempted in the fully relativistic BS approach (but without 
D$\chi$SB and without a dressed $qq\gamma$ vertex), one finds 
that such BS calculations tend to yield the widths
$W(\eta_c\to \gamma\gamma)$ typically below 4 keV -- see 
for example Ref. \cite{Mu"nz96}, containing quite broad fits of 
$\gamma\gamma$-decays of numerous mesons including $\eta_c$,
and references to other related BS calculations, which lead
to similar results. Of course, many of such BS calculations 
could fit
$\eta_c\to \gamma\gamma$ by fine--tuning of model parameters 
aimed especially at $\eta_c$, but simultaneous broad fits of 
many different quantities besides $W(\eta_c\to \gamma\gamma)$ 
performed by Ref. \cite{Mu"nz96} and references therein, indicate 
that the fully relativistic bound-state approaches tend
to reduce the $\eta_c\to \gamma\gamma$ width too far below
its current experimental PDG average \cite{PDG96}. 
On the other hand, our coupled SD-BS approach, concretely
Jain and Munczek model \cite{jain93b} plus GIA, is also 
fully relativistic, but predicts the somewhat higher value
$W(\eta_c\to \gamma\gamma)=5.3$ keV. This is thanks to its 
distinct characteristic -- namely employing also in the 
heavy-quark sector, the quark--photon vertex dressed consistently 
with the dynamical dressing of the quarks (needed for D$\chi$SB 
essential in the light-quark sector) --
which partially compensates the width reduction caused by
the relativistic treatment. Although this is not enough to
approach the current experimental average $W^{PDG}_{'96}$
within one standard deviation, what is in our opinion more
important, is that our result fits
comfortably (and without any adjustment of model parameters
whatsoever!) within the empirical error bars after
the 1996 PDG average is supplemented with the 1995 CLEO 
\cite{CLEO1995} result of $(4.3\pm 1.0\pm 0.7\pm 1.4)$ keV,
resulting in $W^{exp}(\eta_c\to \gamma\gamma)=6.2\pm 1.2$ keV.
In contrast to this, it seems -- to the best of our knowledge --
that unless they fine-tune their parameters for that purpose,
most of the other theoretical approaches predict 
$\eta_c\to \gamma\gamma$ decay widths
that will be  either too large or too small after the
inclusion of the 1995 CLEO \cite{CLEO1995} result in the
PDG average. This is the reason we judge our SD-BS approach
especially successful for $\eta_c\to \gamma\gamma$.

We stress that the 1995 CLEO \cite{CLEO1995} results on 
$\eta_c \to \gamma\gamma$ are reliable and should be taken 
into account \cite{privSavinov}. Admittedly, they have not 
been published in a journal yet (and therefore -- contrary
to our expectations \cite{KeKl1} -- have {\it not}
entered in the current \cite{PDG96} PDG average), but
this is not due to the $\eta_c$ data being questionable. 
It has been due to 
some unresolved issues concerning the $c\bar c$ {\it scalars}
$\chi_{c0}$ and $\chi_{c2}$. 
This has prevented the publication of all $c\bar c$ results of
\cite{CLEO1995}, even though the results for $\eta_{c}$ presented in
\cite{CLEO1995} {\it are} definitive \cite{privSavinov} and in
fact, judging by their error bars, of higher quality than most other
measurements contributing to the present average \cite{PDG96}.
(The inclusion of the $\eta_{c}$-results presented in \cite{CLEO1995},
would reduce the standard deviation of the 1996 PDG average \cite{PDG96}
by some 20\%.)
While the formal reasons for these $\eta_{c}$-results not entering 
in compilations such as \cite{PDG96}, are clear in the absence of 
their publication in a journal, it is also clear under the 
circumstances explained above, that {\it (a)} they should not be 
left out from discussion in research works, and {\it (b)} they
should serve as a particularly strong reminder that new measurements 
and data re-analyses
are still needed in order to make the empirical value of the 
$\eta_{c}\to\gamma\gamma$ width really trustable.

It is important to note that we have not done any
fine--tuning of the parameters, or of the gluon propagator
form (\ref{factorG})-(\ref{gluon_IR}) which we used; these are
the  propagator and the parameters of Ref.~\cite{jain93b}, which
achieved a broad fit to the meson spectrum and pseudoscalar decay 
constants.  Therefore, in the consistently applied generalized 
impulse approximation, the BS approach genuinely, without fitting,
leads to the adequate amplitude strength, which is otherwise too low.
In other words, the BS approach which is in accordance with the ideas 
of \cite{bando94,Roberts,Frank+al}, seems to be able to describe 
electromagnetic processes well even in the heavy-quark sector without 
fine-tuning of model parameters.
On the other hand, since $W(\eta_c\to \gamma\gamma)$ {\it does}
depend on the interaction kernel
and the parameters determining the internal structure of $\eta_c$,
making the measurements of the processes such as 
$\eta_c\to \gamma\gamma$ more 
precise can, through our theoretical approach, contribute to
determining the non-perturbative gluon propagator more accurately.
In the first place, this pertains to
the infrared part, which is at present poorly known,
but is also of significance for the determination
of the QCD running coupling $\alpha_{\rm st}$
(contained in (\ref{gluon_UV}), the ultraviolet part of our gluon
propagator), for example, at the scale $m_c$ naturally sampled by 
the $\eta_c$ system.

To close the circle, the calculations for $\eta_c$ and $\eta_b$
and their $\gamma\gamma$-decays bring us back to the pion, because 
the question arises: why treat the pion in the chiral limit only? 
Why not perform also for $\pi^0$ and its $\gamma\gamma$-decay the 
fully massive calculation?  Well, until recently there was little 
motivation for doing this for the $\pi^0\to\gamma\gamma$-decay,
because it was clear that for this lightest pseudoscalar meson, the 
pure anomaly result (\ref{AnomAmp}) for $T^{chiral}_{\pi^0}(0,0)$ 
is an excellent approximation, and that one can get only very small
corrections to (\ref{AnomAmp}) by discarding (\ref{ChLimSol})
and computing $\pi^0\to\gamma\gamma$ beyond chiral limit.
Recently, however, there appeared another treatment
of $q\bar q$ bound states in the coupled SD-BS formalism
beyond chiral limit \cite{SavkliTabakin97} but also
beyond ladder approximation, which obtained as much as 14\% 
deviation from the pure anomaly amplitude (\ref{AnomAmp}). 
This is not only three standard deviations away from the 
experimental value, but also calls for some clarifications 
as to what corrections to the anomaly result can occur {\it if} 
the massive pion has the pseudo--Goldstone character consistent 
with PCAC.

For the massive pion, the amplitude cannot be obtained analytically 
any more. We have to evaluate $T_P(0,0)$ numerically, just like we 
did for the very massive $\eta_c$ and $\eta_b$ in \cite{KeKl1}. The 
result for the massive pion is, however, little changed with respect 
to the purely anomalous result (\ref{AnomAmp}): numerically,  
$T_\pi(0,0)\approx 0.942 \, T_\pi^{chiral}(0,0)$. 
Rounded to two significant digits, our prediction for the massive 
$\pi^0\to\gamma\gamma$ amplitude is thus just one standard
deviation below the central experimental value -- see Table I.
In the case of the massive pion, there is of course
{\it some} structure dependence, and since the largest influence
on the choice of the model {\it Ansatz} for $G_{IR}$, as well as
on the fixing of the model parameters in \cite{jain93b},
was exercised by mesons much heavier than pions simply because
they are most numerous, it is not surprising that the agreement
with experiment is worsened with respect to the chiral limit where 
the decay amplitude is purely anomalous and thus structure-independent. 
What is important is that in spite of this, the deviation
from the axial anomaly result is not larger than what is allowed
by PCAC. Our amplitude for the massive pion deviates from the purely
anomalous, structure-independent result in the chiral limit by
5.8\%, which still permits the consistency of the used bound-state
model \cite{jain93b} with the Goldstone character of the pion
in the chiral limit and consequently with the PCAC. Namely,
on top of the anomaly, Veltman-Sutherland theorem allows
only contributions which are at least of order $p^2$ higher
than the anomaly result. (For a clear exposition thereof,
see, {\it e.g.}, \cite{Yndurain}). 
This means that the corrections with respect to the anomaly result 
can be of the order of $M_\pi^2/\Lambda_H^2$ \cite{Donoghue+alKnjiga}.
Here, $\Lambda_H$ is a typical hadronic
scale. $M_\pi$ must be an order of magnitude smaller than
$\Lambda_H$ in order that PCAC can make sense.
Such a scale is usually taken to be roughly of the order of
the $\rho$-meson mass $M_\rho \approx 770$ MeV ({\it i.e.}, like
2 constituent quarks forming an ordinary, non-Goldstone meson),
or the scale $\Lambda_\chi$ in the $\chi$PT expansions, given
by ({\it e.g.}, see \cite{L95} and \cite{Chivukula+al93})
\begin{equation}
\Lambda_\chi = \frac {4 \pi f_\pi}{\sqrt{N_{lf}}} \, .
\end{equation}
For the number of light flavors $N_{lf}=3$ [meaning the number of 
quark flavors building up the octet of light pseudoscalar mesons 
that can be regarded as (pseudo)Goldstone bosons], its value is 
$\Lambda_\chi = 670$ MeV. This leads to the rough estimate
that for the realistically massive pion, the deviation from 
the chiral-limit $\pi^0\to\gamma\gamma$ amplitude (\ref{AnomAmp})
cannot exceed much the order of 4\%. 
Our 5.8\% deviation from the anomaly amplitude (\ref{AnomAmp}), 
effectively given by integrals over dressed propagators, the pion 
BS amplitude and dressed $q\gamma q$ Ball-Chiu vertices, is of this 
order and therefore consistent with PCAC and Veltman-Sutherland 
theorem. Since PCAC is really not a hypothesis any more, but is
understood on the basis of the pseudo-Goldstone character
of the pion, our massive $\pi^0\to\gamma\gamma$ result is
a check of the pion being -- in the Jain-Munczek model 
\cite{jain93b} -- both a $q\bar q$ bound state {\it and} 
the proper pseudo-Goldstone particle.
In contradistinction to that, the 14\%-deviation from
the anomalous amplitude $T_\pi^{chiral}(0,0)$, obtained by
\cite{SavkliTabakin97} for a massive pion, would come from
$\Lambda_H=360$ MeV $=2.6 M_\pi$. This is larger than
$M_\pi$, but not by an order of magnitude. Such a
bound-state model therefore comes in the conflict
with PCAC. This is an indication that their pion is
not a proper pseudo-Goldstone particle. In fact, the 
approach of Ref. \cite{SavkliTabakin97} is a specific
attempt to go beyond the ladder approximation for the 
interaction used in the SD and BS equations. We take 
their result on $\pi^0\to\gamma\gamma$
as an indication  that their SD and BS equations are
not coupled into a consistent approximation scheme any
more. It is well-known that (rainbow-)ladder approximation --
as used in \cite{jain93b} for example -- is such a consistent
approximation scheme, which preserves the Goldstone pseudoscalar
in the chiral limit, and the pseudo-Goldstone one for quarks
which are light in comparison with $\Lambda_{QCD}$.
It is very desirable to go beyond this approximation,
but it is obviously not yet clear how to do it and
preserve successes of the coupled SD-BS regarding the
axial anomaly and pseudo-Goldstone character of the
pion.

\section{Concluding discussion of the model}

The concrete realization of the coupled SD--BS approach used 
in this work, namely the model of Jain and Munczek, is quite
successful from the lightest to the heaviest meson masses. 
It should thus have wide future applicability, 
especially because this feature is combined with another one 
(common to coupled SD--BS approaches in general), namely that 
it is manifestly relativistically covariant. One can therefore
relate its solutions -- and thus also form factors, {\it etc.} --
in different frames, for high recoil momenta, which becomes 
increasingly important with the advent of ever more energetic 
facilities (for example TJNAF today, tomorrow ELFE, {\it etc.}).

Therefore, we would like to close this paper with a 
discussion of the model of Jain and Munczek, with the emphasis on
clarifying how this model can -- over such a broad range of 
masses -- be so successful in describing the meson spectrum and 
pseudoscalar decay constants \cite{jain91,munczek92,jain93b},
and, as shown in \cite{KeKl1} and this paper, also in
describing the $\gamma\gamma$-interactions of pseudoscalars
(including the anomalous $\pi^0\to\gamma\gamma$),
especially if one notes that all this was done 
with a small number of parameters.

This clarification is necessary, because it is known that
the sector of light quark mesons is rather different from the
one of heavy quark mesons, and at first is hard to understand
how it is possible to treat them in such an unified manner as
in the model of Munczek and Jain \cite{jain93b}. We especially
want to clarify how is this possible in spite of employing: 
{\it i)} the ladder approximation, which has been almost 
universally used to make BS calculations tractable, but 
which is an uncontrolled approximation after all, and 
{\it ii)} a model assumption, namely the model {\it Ansatz} for 
$G_{IR}$, which has been justified only {\it a posteriori}, by its 
phenomenological success.

In the following, we will basically argue that good results 
over the wide mass range are 
indeed not accidental and are in fact understandable, because the 
way this model has been constructed suggests that it had the 
potential to combine (and hence in some things surpass)
the successes of two reputable models: A.) Nambu--Jona-Lasinio (NJL)
model which is successful in the light sector because it incorporates
D$\chi$SB, and B.) the constituent quark model which is certainly 
adequate for heavy quarks.

If one notes that Jain and Munczek's approach includes the 
merits of both the NJL model and the constituent quark model,
the success of Jain and Munczek's model \cite{jain93b} in describing
the meson spectrum from light to heavy is not so mysterious.
Let us remember that various non-relativistic and relativistic
constituent quark models work quite well also for {\it almost} 
entire spectrum of hadrons -- not only mesons, but also baryons. 
However, they do {\it not} perform well for light pseudoscalar 
mesons even if we talk just about the spectrum. Besides the
references already discussed in Section 2, let us
recall another work on meson spectroscopy of Sommerer 
{\it et al.} \cite{Somm+al} as an illustrative example.
In this well-known paper, the spectrum of almost 50 meson 
states was fitted in a relativistic
constituent quark model. Their interaction kernel, in the
ladder approximation and same for all fitted mesons, consists
of a one-gluon exchange interaction (roughly analogous to
our $G_{UV}$) and a phenomenological, long range ({\it i.e.},
roughly analogous to our $G_{IR}$) ``string tension" potential.
They do not have D$\chi$SB, and constituent quark masses are
simply parameters. Their so--called ``reduction B" (their assumed
{\it prescription} for reducing their BS equation from 4 to 3 
dimensions) results in successful reproduction
of even the very lightest, $\pi$--mass. However, 
they themselves note that there are symptoms that 
not all is well for small masses, because their other
prescription for reducing their BS equation from 4 to 3 dimensions,
``reduction A", does poorly at representing these low--mass mesons,
even though there is no apparent physical reason why one of those
{\it assumed} prescriptions should be favored over the other. 

Therefore, this is another interesting case showing that one can expect
troubles for small masses if one treats them too similarly
to the large ones; however, our point of view is that what is missing
in the works of this kind
is primarily D$\chi$SB, which determines the properties of pseudoscalars
in the light quark sector. When D$\chi$SB is not included, there will
be a price to be paid in the light sector: even if troubles with the
light spectrum are avoided by fine--tuning model parameters and opting 
for most favorable reduction prescriptions, there is no escape from the
trouble with the anomalous processes such as $\pi^0\to\gamma\gamma$.

On the other hand, Jain and Munczek's model is constructed to
have the correct limiting behavior {\it both} in the non-relativistic 
limit (for heavy quarks and antiquarks of the same flavor), {\it and}
in the chiral limit. Indeed, we will argue below that the success of
Jain and Munczek's approach in the light-quark sector is tied more to 
the incorporation of D$\chi$SB than to any of the specific interaction 
kernels used \cite{jain91,munczek92,jain93b}. As for the heavy quark 
sector, even if their model cannot in all applications compete -- in 
that sector -- with a specialized scheme such as Heavy Quark Effective 
Theory (HQET \cite{Neubert94}, which explicitly takes advantage of 
simplifications available in the heavy-quark region), it is certainly 
capable of providing a bridge between these two limiting regions. As 
stressed by Burden \cite{Burden98}, the understanding of nonperturbative 
dynamical self-dressing -- which is the key feature both in QCD
and in the SD-BS approach to QCD -- is important also for the heavy 
quarks and for the proper understanding of the successes of HQET. 
The introductory section of Munczek and Jain's second paper 
\cite{munczek92} is especially detailed in discussing their
motivation (and to some extent, even justification) for resorting to 
the ladder approximation as a -- for the time being -- unavoidable
but also acceptable modeling assumption, so here we just point out
that the {\it ladder} BS equation is known (\cite{FlammSch82} and
references therein) to reproduce the Schr\" odinger equation in the 
non-relativistic limit, namely when both quark masses become very 
large. In fact, Munczek and Jain \cite{munczek92,jain93b} have even 
derived the non-relativistic potentials following from the
gluon propagators assumed, and they are in reasonable
agreement with the potentials used in non-relativistic models.
(The limitations of the ladder approximation do lead to additional
difficulties when one of the masses is much larger than the other.
Later we will comment on this some more, but now let us remember that
in the present paper we have dealt exclusively with  quarks and antiquarks
of the same flavor.)

So, the model of Munczek and Jain is certainly no worse 
than the non-relativistic quark model or the relativistic 
constituent quark model.
In fact, for light quarks it is better than these other
models because it satisfies chiral symmetry, enabling
the correct chiral limit behavior,
complete with the appearance of Goldstone bosons due to D$\chi$SB,
essentially in the Nambu--Jona-Lasinio fashion \cite{NJLa,NJLbc}.
Thus all the results obtainable from pure symmetry arguments
are automatically satisfied in this model. This is very
satisfying and, as already stressed above, will happen only
if the consistency between SD equation and BS equation is
maintained. However, this consistency does not pertain just
to the usage of the same interaction, that is the same gluon
propagator; SD and BS equations must
be treated in the consistent approximations.
In particular, since the work of
Nambu and Jona-Lasinio \cite{NJLa,NJLbc},
the Goldstone boson solution of the BS equation was
found to accompany the spontaneous chiral symmetry breaking
appearing in the SD equation for the quark propagator
in a variety of chirally invariant models; however, 
let us recall that until recently (see Ref. \cite{Munczek95}), 
this result was established only
when {\it both} SD and BS equations were in the
ladder approximation. (In the case of SD equations, this
approximation is also often called ``the rainbow 
approximation{\footnote{In a more precise terminology of references 
such as \cite{Roberts,RW,Burden+Qian+al}, a framework  such as
the coupled SD-BS Jain--Munczek approach is called
the ``rainbow--ladder framework" or  the ``rainbow--ladder
truncation of the quark DS equation and two--body
BS equation". However, in this paper, we follow somewhat 
less precise, but shorter Jain and Munczek's terminology.}}").
Of course, it would be very desirable
to go beyond the ladder approximation, 
and in fact in his paper on dynamical chiral symmetry breaking, 
Goldstone's theorem, and the consistency of the SD and BS equations,
(which we just quoted above) Munczek himself \cite{Munczek95}
discusses in which way the BS equation should change, after the 
SD equation has been taken beyond the ladder approximation, in
order to preserve the appearance of Goldstone bosons
and other dynamically broken chiral symmetry features.
He has made progress, as have also some other authors 
\cite{Bender+al96}, but the implementation in a concrete
coupled SD-BS bound state model which would be elaborated analogously
to the present Jain-Munczek model, {\it and} which would go beyond
the ladder approximation, is still not an immediate possibility,
and must be left for some later stage.

To summarize this issue:

{\it i)} The ladder approximation does give
the correct non-relativistic limit.
Jain and Munczek's model therefore also has the
the correct non-relativistic limit, so that there
are no problems for systems where both quarks
are heavy. (As Ref. \cite{munczek92} cautioned, 
ladder approximation might cause problems when one of 
the quark masses becomes much larger -- {\it i.e.},
in heavy--light systems, but we do not treat them in 
the present paper. Also, see the comment in the last paragraph.)

{\it ii)} On the
other hand, the applicability of the ladder approximation to
the light systems is not really known, and the work of  Munczek
and Jain is an attempt to investigate that. Their results,
along with some Miransky's arguments (see especially the
references of Miransky and his collaborators in \cite{munczek92})
and the recent results of Maris, Roberts and Tandy
\cite{MarisRoberts97PRC56,MarisRobertsTandy97},
are quite encouraging in this respect.

{\it iii)} Dynamical chiral symmetry breaking can be incorporated
correctly (complete with the appearance of Goldstone pseudoscalars in
the chiral limit) in the SD--BS framework using the ladder approximation.
(In fact, until recently it was the only way one knew how to do it,
and it may still remain one of the most practical ways for modeling
even in the future. For example, Miransky \cite{Miransky} in his Chapter 12
judges that such models based on the improved ladder approximation
are an improvement with respect to the widely used Extended
Nambu--Jona-Lasinio model.)

Keeping in mind {\it i)} and {\it iii)},
it is rather transparent how it was possible for Jain and Munczek
to construct a model where the results
for the light $q\bar q$ ground-state masses were found to
be in good qualitative and quantitative agreement with
current algebra results, while for
heavy quarks such as $b$ and $c$ the predicted masses of their bound
states are proportional to the sum of their ``constituent"
masses in accordance with nonrelativistic limit expectations \cite{jain93b}.

As for the point {\it ii)}, the numerous empirical successes of Jain 
and Munczek's model, especially for pseudoscalar and vector mesons (whereas
the light scalars remain problematic), have provided a rather strong 
{\it a posteriori}
indication that the ladder approximation is reasonably useful for the
light quarkonia. In addition to that, recent investigations beyond the
ladder approximation seem to provide new insight why it is so. It turns
out \cite{Bender+al96,RobertsProceed97}  
that among the corrections beyond ladder approximation that have been
computed so far, those computed for pseudoscalar, vector and axial $q\bar q$
bound states tend to roughly {\it cancel out}, whereas for scalar and tensor 
$q\bar q$ bound states they mostly {\it add up}. If this trend eventually 
turns out to continue also for higher corrections in the pseudoscalar, vector
and/or axial channels, this will show that the ladder approximation is indeed
reasonable for these channels. Even if one objects that so far this is still
not a proof, and that Jain and Munczek \cite{jain93b} maybe had just an
accidental success in the above point {\it ii)}, or, even if one
insists that one has to go beyond the ladder approximation in the
light sector to get a good description of the internal hadronic
structure there, the point {\it iii)} still permits Jain and Munczek's
model to unquestionably retain all good results in the light sector
that follow just from symmetry and the correct incorporation of
the dynamical chiral symmetry breaking.

In that connection, recall the following: of the $\gamma\gamma$-processes
we calculated in the light sector, the chiral $\pi^0\to\gamma\gamma$
is not affected at all, and $\gamma^\star\pi^0\to\gamma\gamma$ as well 
as the massive $\pi^0\to\gamma\gamma$ are affected rather little by how 
Jain and Munczek's
model describes the internal pion structure, and here is why.
We have found (along with, {\it e.g.}, Bando {\it et al.} \cite{bando94}
and Roberts \cite{Roberts}) that, in the chiral ({\it and} soft) limit,
our amplitude for $\pi^0\to\gamma\gamma$ is completely independent of 
the bound-state vertex for the Goldstone boson case 
($\propto \gamma_5 B(q^2)$) {\it I.e.}, the pion structure falls out 
completely for the $\gamma\gamma$ decay of Goldstone pions, as it should, 
because there the axial anomaly dominates. In other words, for our 
treatment of the massless $\pi^0\to\gamma\gamma$ the only important 
thing was the correct incorporation of the symmetry properties: the
dynamical chiral symmetry breaking leading to Goldstone bosons
in Jain and Munczek's model,
and the Ward-Takahashi identity of QED respected by GIA!
It did not matter exactly what
solution $B(q^2)$ we got, and from what interaction kernel. We
analytically obtained exactly the same answer (the correct anomaly
amplitude) as Roberts \cite{Roberts} 
with his {\it Ansatz} for $B(q^2)$, even though it was
very different from our chiral--limit solution for $B(q^2)$ -- and as
just said, it was because the internal structure
did not matter -- only obeying the Ward-Takahashi identity
and the correct chiral limit behavior mattered.
(A place where we can, and in fact do,
detect some structure dependence is
the off--shell extension $\gamma^\star\pi^0\to\gamma$.
However, it seems that this dependence is rather weak, since we
get results relatively similar to Frank {\it et al.} \cite{Frank+al}, 
even though our solution for $B(q^2)$ is very different from their
{\it Ansatz} for $B(q^2)$. In $\pi^0\to\gamma\gamma$ away from the
chiral limit there is also some room for structure dependence, but
again on the level of several percent, in keeping with PCAC and 
Veltman--Sutherland theorem. The correct incorporation of D$\chi$SB
obviously remains the dominant feature.) 
Therefore, even if a sceptical reader
still doubts that Jain and Munczek's model can simultaneously give
realistic descriptions of the hadronic structure for
both heavy and light systems, our simultaneously good results
for both $\pi^0\to\gamma\gamma$ and $\eta_c\to\gamma\gamma$
need not be suspect. It is in fact sufficient that: 1.) the
GIA with the Ball--Chiu vertex is a good way to incorporate
the interactions of dressed quarks with photons,
and that 2.) we have a reasonable description
of the hadronic structure for $\eta_c$ and $\eta_b$, because 
what matters for $\pi^0\to\gamma\gamma$ is D$\chi$SB and 
the (pseudo)Goldstone character of $\pi$, 
and not details of the internal pion structure.

With that we finish our detailed explanation of functioning of 
our chosen variant \cite{jain91,munczek92,jain93b} of the 
coupled SD-BS approach. We close the article with some additional
comments on the situations in the presence of heavy quarks where 
the present approach is useful and on the situations 
where, due to the limitations of the present approach such as the 
ladder approximation, more accurate results are expected from
simpler approaches using the non-relativistic approximation.  

Possible simplifications are always something valuable. 
In particular, it is clear that there is no way of beating 
the accuracy of non-relativistic
approximation in the limit of very heavy quark mass.
However, the question is what is the value of that quark mass.
For top this is certainly true. Bottom is also treatable
very well by non-relativistic approximation. In the
case of $\eta_b$, we do not mean to challenge the
accuracy of the nonrelativistic description; nevertheless,
it would
be instructive to see how well (or not so well) the coupled
SD-BS approach does there
once the experimental data on $\eta_b$ are obtained.
However, for $\eta_c$ we believe that relativistic effects
and effects of dynamical dressing on the quark-photon processes
can be very important -- especially in the light of recent experimental 
results from CLEO \cite{CLEO1995}. Besides already quoted Ref. \cite{Mu"nz96}, 
which advocates the view that relativistic effects are important
for all $\gamma\gamma$ widths, including heavy quarkonia, let us
also give the corroborating views of {\it (i)} Z\" oller {\it et al.}
\cite{Zoller+al} who found that relativistic effects are 
non--negligible even for mesons involving only heavy quarks and 
in fact particularly important for $M1$ transitions in charmonium, 
then of {\it (ii)} Resag and M\" unz \cite{RM95}, who point out that  
in charmonium one still finds typical velocities of $v/c \approx 0.4$ 
\cite{HardSuch85}, so that relativistic effects should become important 
especially for electroweak decay properties, as has been shown by 
Beyer {\it et al.} \cite{Beyer+al92}.  Also, we remark that for the 
hidden flavor systems, where constituents have equal masses,
there were no problems in obtaining accurate solutions.
On the other hand, in the cases such as $B$--mesons, where one quark 
is significantly heavier than the other, we did have problems in 
reproducing Jain and Munczek's solutions as accurately as in the 
equal-mass cases.  This is not surprising, because it is known that the 
ladder approximation is especially troublesome for such cases, as Jain 
and Munczek \cite{munczek92} themselves warned. 
Obviously, these are the cases where we may profit from making a systematic 
nonrelativistic reduction, say by expanding in powers of the inverse mass 
of the heavier quark and adopting elements of HQET, as well as utilizing
the insights of SD--BS studies pertinent for heavy--light systems, such 
as \cite{Ivanov+al98PLB,Burden98,Ivanov+al98PRC}.
Such considerations become really important when heavy--light 
systems are addressed.  However, in the case at hand, for the 
systems built of a quark and its equally massive antiquark, the 
straightforward usage of accurately reproduced solutions of Ref. 
\cite{jain93b}, is adequate in every respect.


\section*{Acknowledgments}
\noindent The authors acknowledge the support of the 
Croatian Ministry of Science and Technology contracts 
1--19--222 and 009802.


\begin{table}
\begin{tabular}{ccccccc}
    &  &  &  &  &  &  \\
$P$ &  $M_P^{exp}$ & $M_P$ & 
${\displaystyle{B(0)}}\over{\displaystyle{A(0)}}$
& $T_P(0,0)$ & $T_P^{exp}(0,0)$ &
${\displaystyle{T_P(0,0)}
}\over{\displaystyle{T^{{chiral}}_{\pi^0}(0,0)}}$\\
\hline
    &  &  &  &  &  &  \\
${\mbox{chiral}}$ $\pi^0$ &   & 0 & 356 & $0.272\cdot 10^{-3}$ &  &  1  \\
    &  &  &  &  &  &  \\
 $\pi^0$   & 135 & 140 & 375 & $0.256\cdot 10^{-3}$ & $(0.27\pm0.01)\cdot 10^{-3}$ & 0.942    \\
    &  &  &  &  &  &  \\
$\eta_{s}$ &     &   721  & 610 & $0.0798\cdot 10^{-3}$ &   & 0.294     \\
    &  &  &  &  &  &  \\
$\eta_{c}$  & 2979 & 2875 & 1540&$0.0692\cdot 10^{-3}$ & $(0.075\pm 0.007)\cdot10^{-3}$  & 0.255 \\
    &  &  &  &  &  &  \\
$\eta_{b}$      &  ?  & 9463      & 4770  &$2.057\cdot 10^{-6}$&  ?  & $7.57\cdot 10^{-3}$ \\
    &  &  &  &  &  &  \\
\end{tabular}
\caption{For the presently considered neutral pseudoscalars $P$
[$\pi^0$ both in the chiral limit and in the realistically massive 
case, the unphysical $s\bar s$-pseudoscalar ($\eta_s$), $\eta_c$ 
and $\eta_b$], we compare their experimental masses, where applicable 
or available, with our respective theoretical predictions.
The ratios $B(0)/A(0)$ roughly represent the constituent quark masses 
for the quark flavors [massless and massive (but isosymmetric) $u$ and 
$d$, as well as $s$, $c$, $b$]  pertinent for these mesons $P$.
The next two columns provide the comparison of the calculated 
$P\to\gamma\gamma$ decay amplitudes $T_P(0,0)$ with their respective 
average experimental widths, where we have included the 1995 CLEO result 
\protect\cite{CLEO1995} on $\eta_c\to\gamma\gamma$.
At present, there are no experimental data on $\eta_b$.
The masses [including the ratios $B(0)/A(0)$] are given in MeV, the 
decay amplitudes in MeV$^{-1}$, while the amplitude ratios are of course 
dimensionless. The model value of the pion decay constant we predict
[and use in the expression (\ref{AnomAmp}) for $T^{{chiral}}_{\pi^0}$]
is $f_\pi=93.2$ MeV.}
\end{table}


\newpage

\section*{Figure captions}

\begin{itemize}

\item[{\bf Fig.~1:}] The UV part (the dashed line) and the 
IR part (the dotted line) of the total presently used effective 
propagator function (the solid line).

\item[{\bf Fig.~2:}] The comparison of the  presently used effective
propagator function (the solid line) with the effective gluon propagator 
functions of \cite{CahillGunner95} (the short-dashed line), of 
\cite{CahillGunner97preprint} (the dash--dotted line) and of
\cite{SavkliTabakin97} (the long-dashed line).

\item[{\bf Fig.~3:}] 
Our chiral-limit solution (the solid line) for the propagator 
function $A(q^2)$ is compared with our massive solutions for
various ${\widetilde m}(\Lambda)\neq 0$ (the dotted lines marked 
by letters denoting the pertinent flavors).
The dashed line denotes the $A(q^2)$-{\it Ansatz}
(for $u,d$-quarks) of \cite{Roberts}, and also of Frank 
{\it et al.} \cite{Frank+al} who have such parameters that
the difference with respect to the dashed line \cite{Roberts}
cannot be seen on this figure.

\item[{\bf Fig.~4:}] 
The comparison of our chiral-limit solution (the solid line)
for the propagator function $B(q^2)$ with our massive solutions
for various ${\widetilde m}(\Lambda)\neq 0$ represented by the 
dotted lines marked by letters denoting the pertinent flavors,
and with the {\it Ansatz} (for $u,d$-quarks) for $B(q^2)$ employed by   
\cite{Roberts} (the dashed line), and that of \cite{Frank+al},
which cannot be distinguished from the dashed line in this plot.

\item[{\bf Fig.~5:}] 
The solid line denotes our constituent quark mass function 
$B(q^2)/A(q^2)$ in the chiral limit, while the dotted lines 
(marked by letters indicating the pertinent flavors)
denote our constituent quark mass functions for
${\widetilde m}(\Lambda)\neq 0$. The one following
from the {\it Ans\" atze} of \cite{Roberts,Frank+al} 
is denoted by the dashed line.

\item[{\bf Fig.~6:}] 
Diagram for $\pi^0,\eta_{c}, \eta_{b} \to\gamma\gamma$ decays, and for 
the $\gamma^\star\pi^0\to\gamma$ process if $k^{\prime 2}\neq 0$.

\item[{\bf Fig.~7:}]
$\gamma^\star\pi^0\to\gamma$ form factor. Experimental points are 
the results of the CELLO collaboration \protect\cite{behrend91}.
The solid line represents our results, while the dashed line represents 
those of Ref.~\protect\cite{Frank+al}. The dash--dotted line is the 
Brodsky--Lepage interpolation \protect\cite{BrodskyLepage}, the line 
of open circles is the curve of \cite{BH}, and open squares form the 
monopole curve corresponding to 
${\langle r^2_{\gamma\pi^0\gamma} \rangle}^{1/2}=0.65$ fm.

\end{itemize}


\newpage

\vspace*{4cm}
\epsfxsize = 15 cm \epsfbox{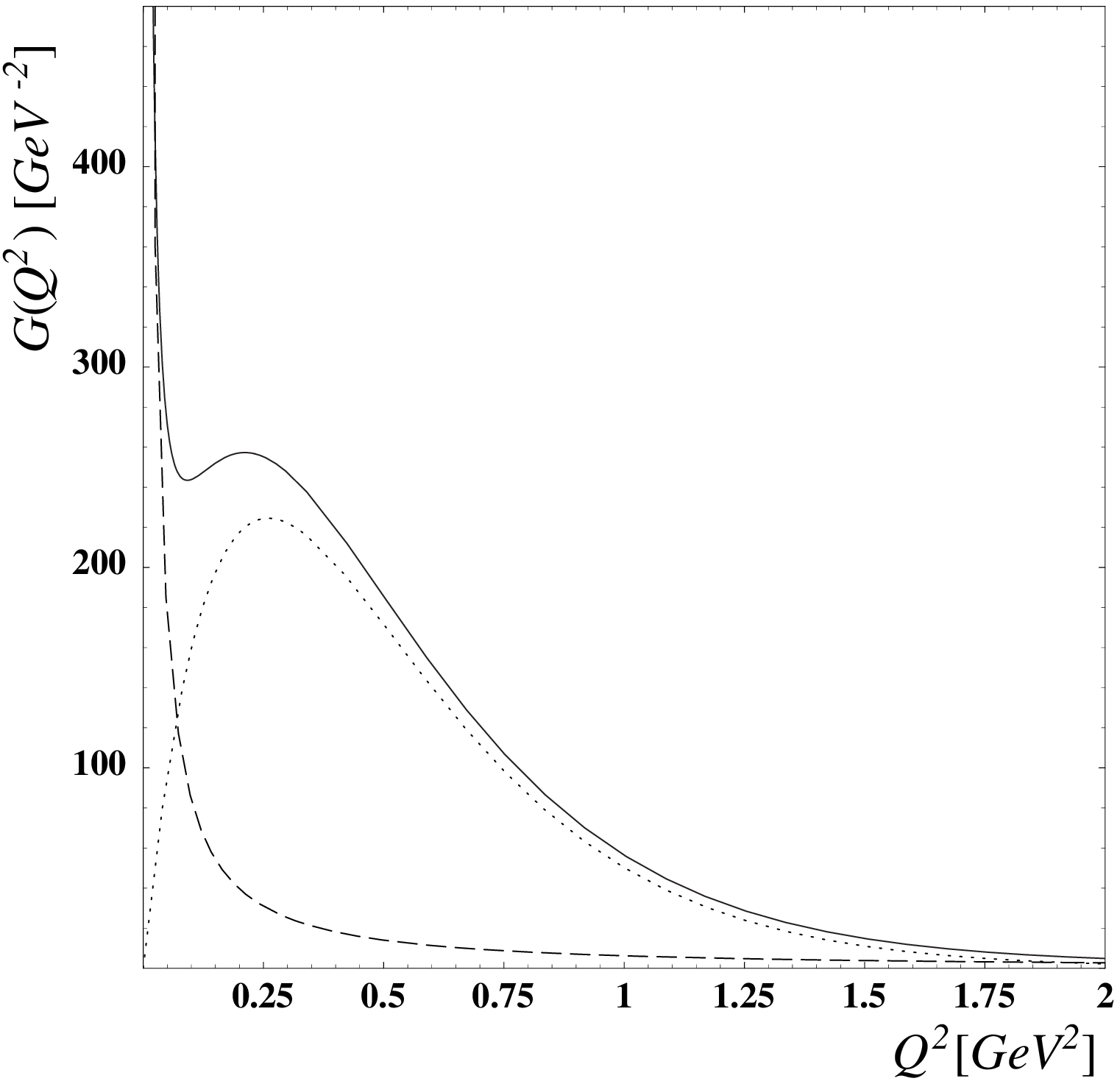}

\newpage

\vspace*{4cm}
\epsfxsize = 15cm \epsfbox{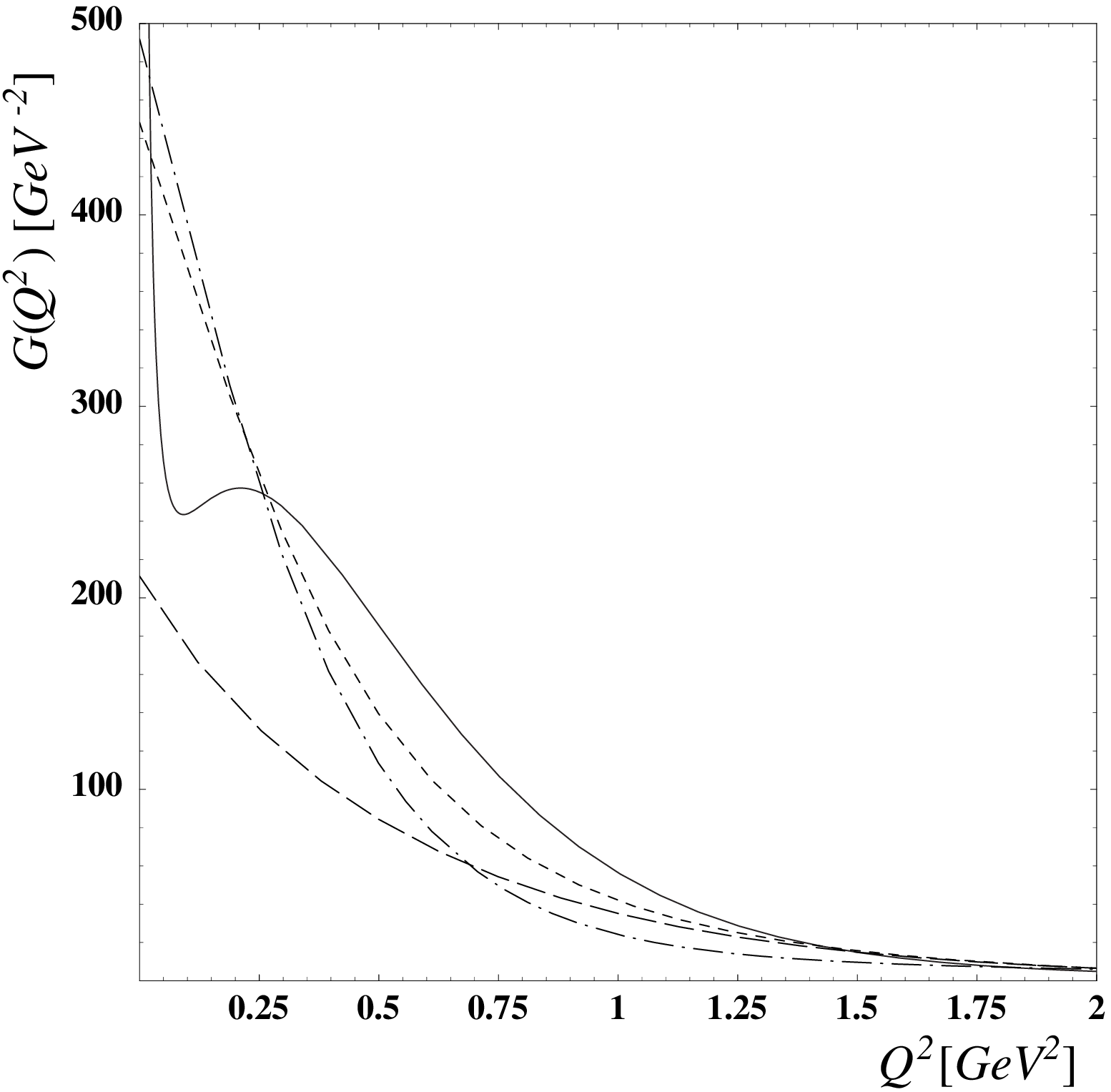}

\newpage

\vspace*{4cm}
\epsfxsize = 15cm \epsfbox{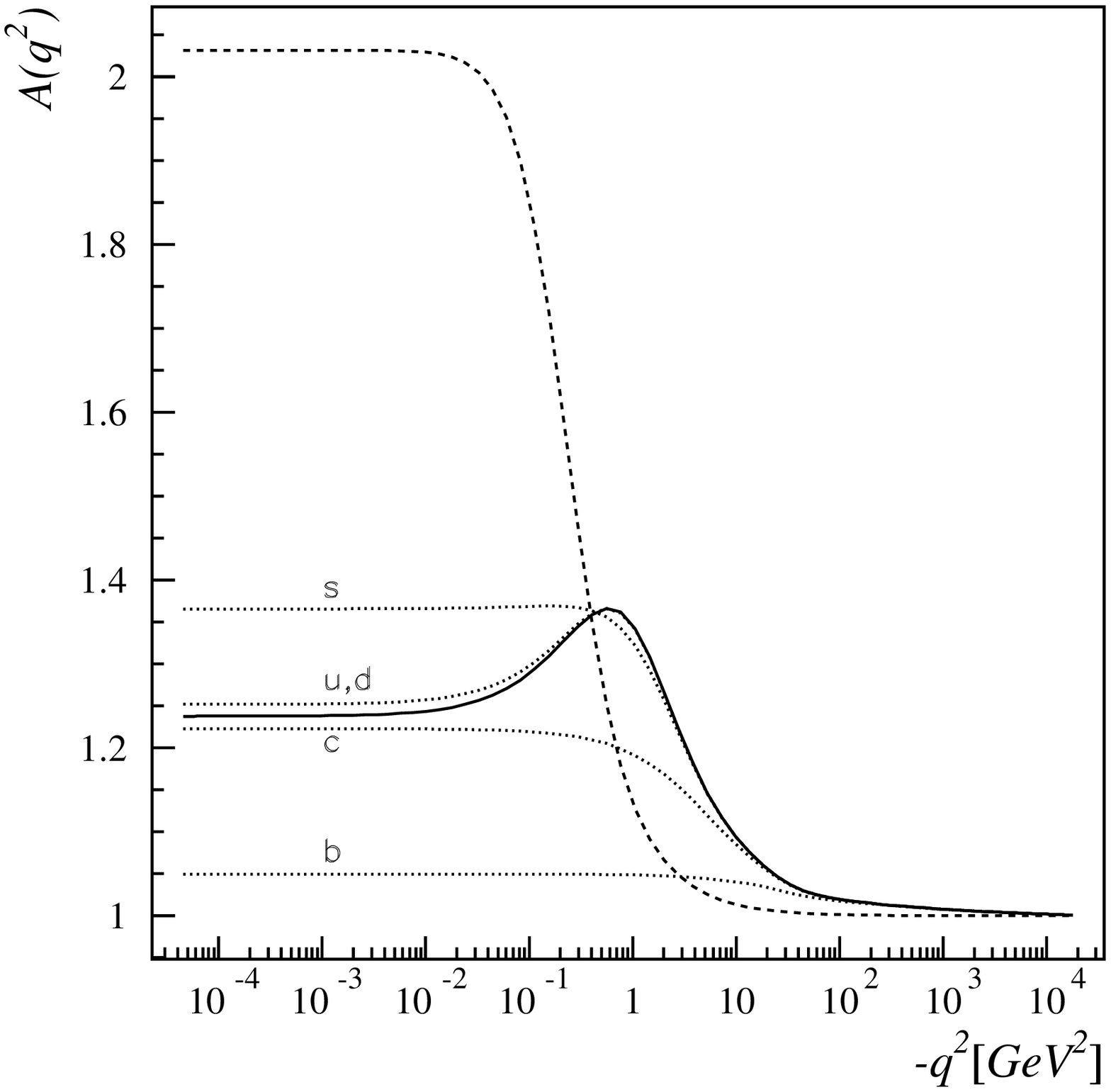}

\newpage

\vspace*{4cm}
\epsfxsize = 15cm \epsfbox{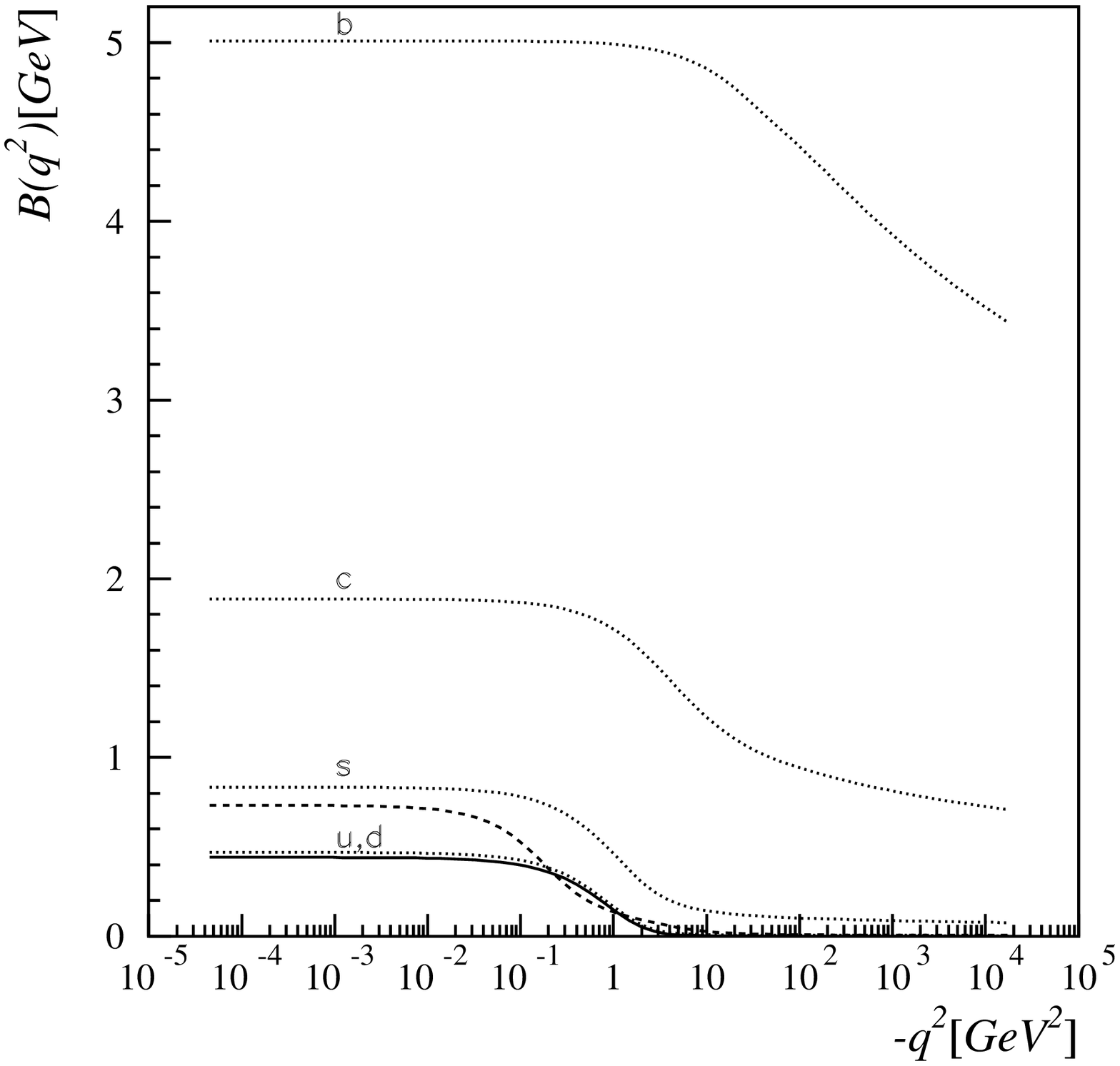}

\newpage
\vspace*{4cm}
\epsfxsize = 15cm \epsfbox{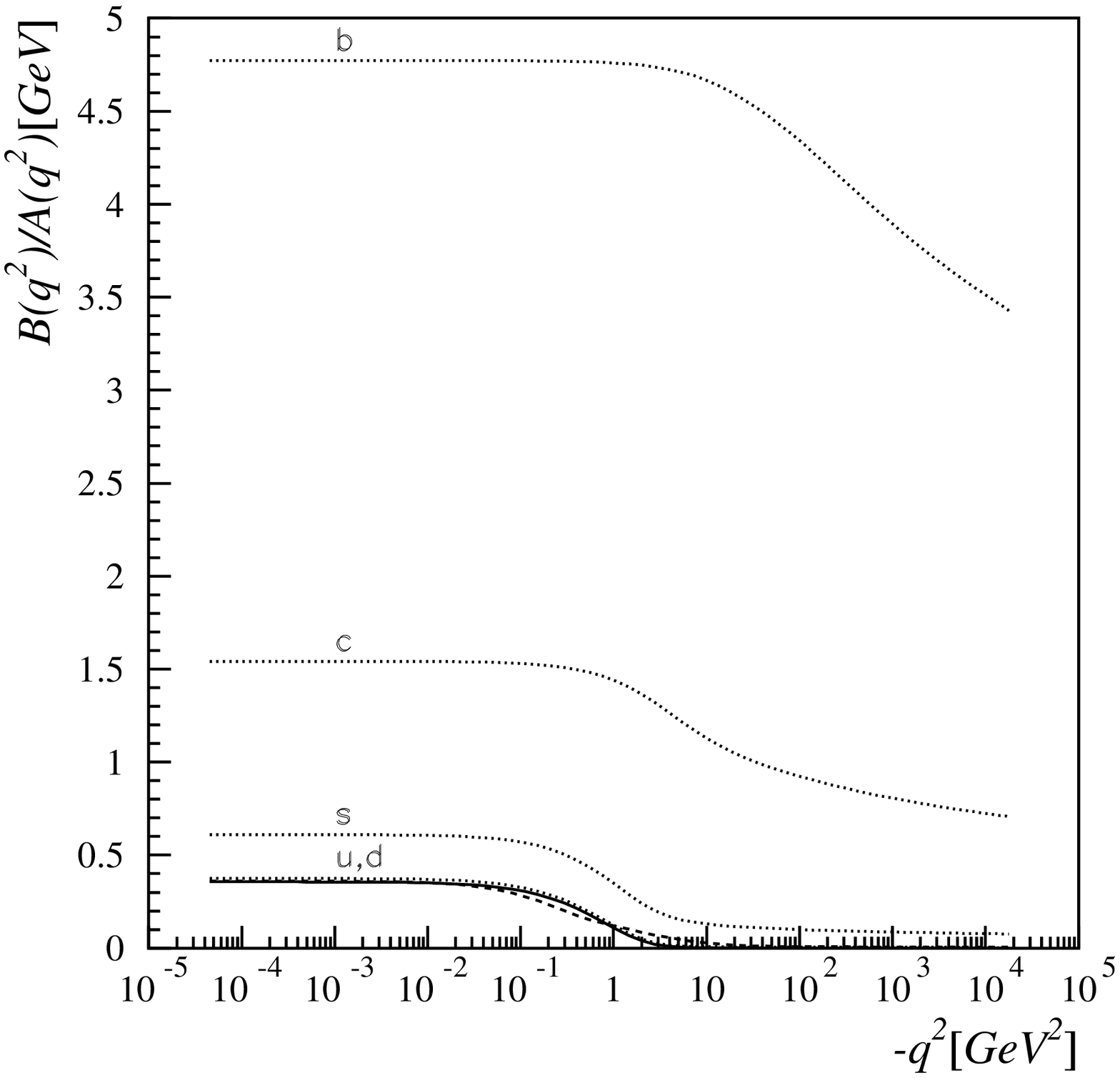}

\newpage

\vspace*{4cm}
\epsfxsize = 15cm \epsfbox{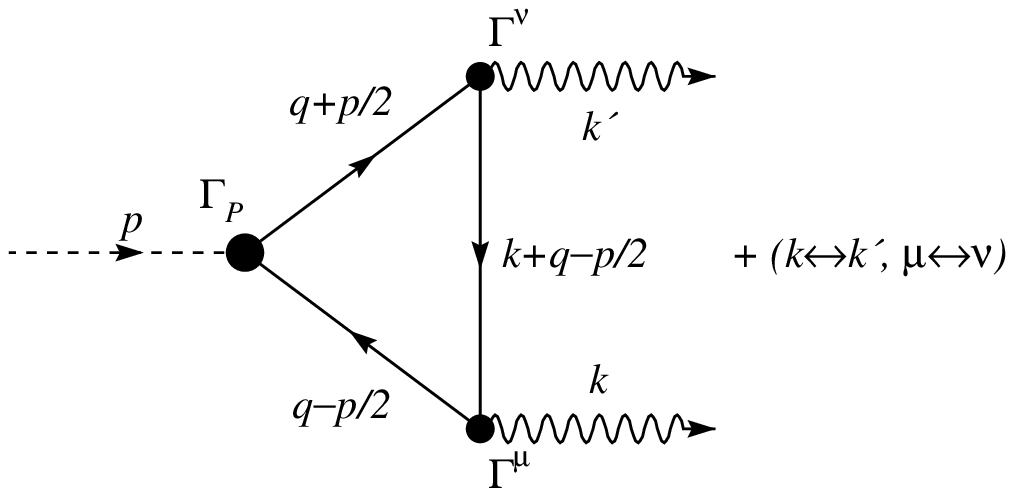}

\newpage

\vspace*{2cm}
\epsfxsize = 15cm \epsfbox{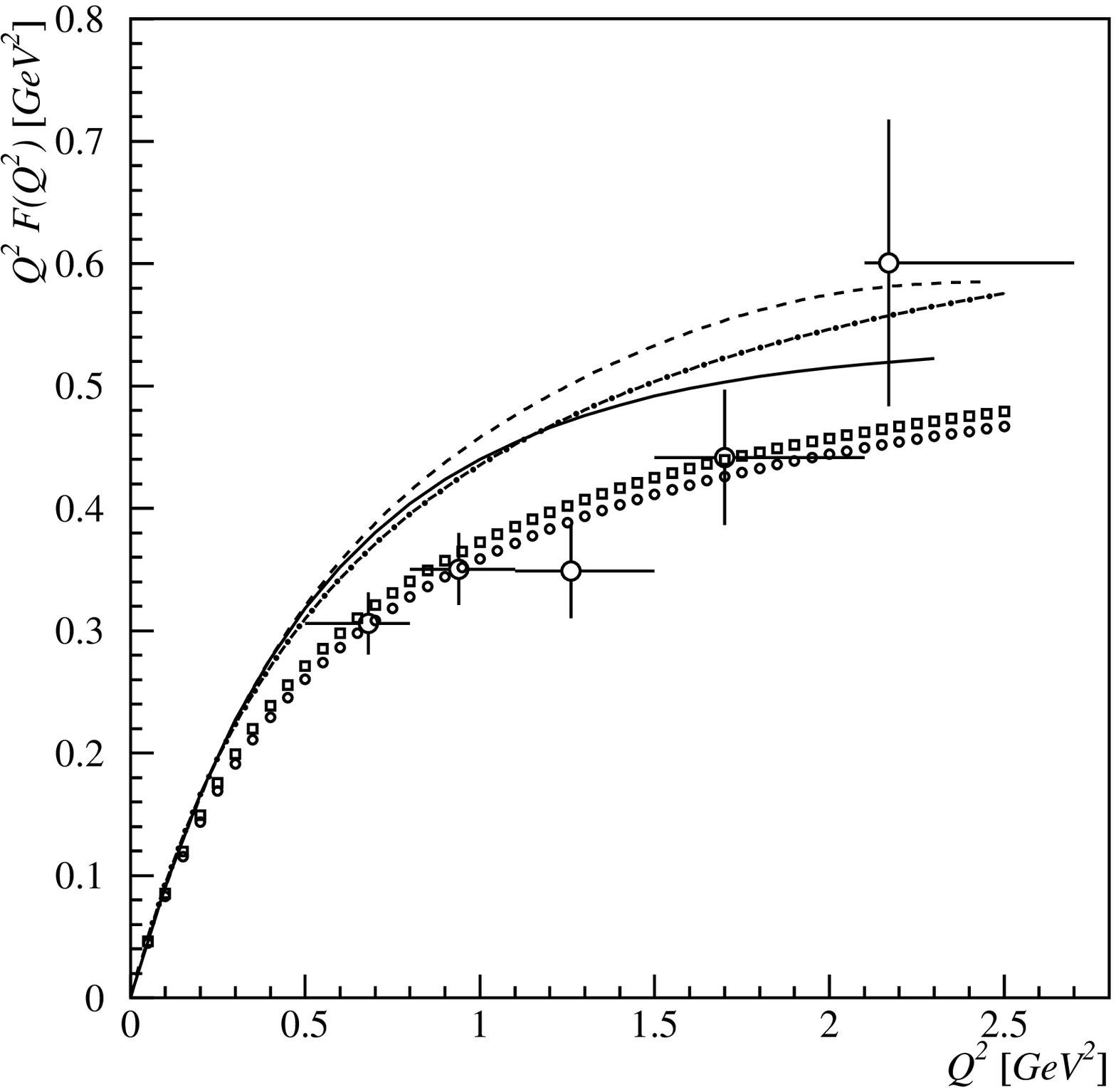}

\end{document}